\documentstyle[galley,epsf]{mn}
\title{ CH stars at High Galactic Latitudes} 
\title[CH stars at High Galactic Latitudes]{CH stars at High Galactic Latitudes}

\author[Aruna Goswami ]{Aruna Goswami$^{ }$\thanks{E-mail:
aruna@iiap.res.in } \\ 
$^{ }$Indian Institute of Astrophysics, Koramangala, Bangalore 560034, India}

\begin{document}

\date{Accepted 2005 Feb 9:  Received  2005 Feb 3; in original form 2004 Nov 24}

\pagerange{\pageref{firstpage}--\pageref{lastpage}} \pubyear{2005}

\maketitle

\label{firstpage}

\begin{abstract}
Carbon-rich stars of population II, such as CH stars, can provide direct
information on the role of low to intermediate-mass stars of the halo on
the early Galactic evolution. Thus an accurate knowledge of CH stellar
population is a critical requirement for building up scenarios for early
Galactic chemical evolution. In the present work we report on   several CH 
stars  identified in a   sample of Faint High Latitude Carbon stars from 
Hamburg survey and  discuss their medium resolution spectra covering a 
wavelength range 4000 - 6800 \AA\,.  Estimation of the depths of bands 
(1,0) $^{12}$C$^{12}$C ${\lambda}$4737 and
(1,0) $^{12}$C$^{13}$C ${\lambda}$4744 in these stars indicate isotopic
ratio $^{12}$C/$^{13}$C ${\sim}$ 3, except for a few exceptions; these 
ratios are consistent with existing theories of CH stars evolution. 
The stars of Hamburg survey, a total of 403 objects were reported to be 
carbon star candidates  with strong C$_{2}$ and CN molecular bands.  
 In the first phase of observation, we have acquired spectra of  ninety 
one  objects. Inspection of the objects spectra show fifty one  objects 
with  C$_{2}$ molecular bands in their  spectra of which thirteen  stars
have low flux below about 4300 \AA\,. Twenty five objects show weak or 
moderate CH and CN bands ,
twelve objects show weak but detectable CH bands in their spectra and 
there are three objects which do not show any molecular bands due to C$_{2}$, 
CN or  CH in their spectra. 
Objects  with  C$_{2}$ molecular bands and with  good signals
bluewards of 4300 \AA\,  which show prominent CH bands in their spectra  are
potential candidate CH stars. Thirty five such candidates  
are found in the present sample  of ninty one objects  observed so far.
The set of CH stars identified could be the 
targets of subsequent observation at high resolution for a detail and 
comprehensive analysis for understanding their role in early Galactic 
chemical evolution.   

\end{abstract}


\section{Introduction}

Knowledge of stellar population offers a fossil record of formation and
evolution of galaxies and thus provide strong constraints on the scenarios
of the Galaxy formation and evolution. Carbon stars, for instance,  were
thought to be giants without exceptions and sought as tracers of the outer 
halo. Recent surveys on stellar populations have led to the discovery of
different types of stars, numerous metal-poor stars, carbon and carbon-related
objects etc. (Beers et al. 1992, Totten and Irwin 1998, Beers 1999).
One of the  results of these efforts is the 
great discovery that  the fraction of carbon-rich stars
increases with decreasing metallicity (Rossi, Beers and Sneden 1999).
Extensive analysis of many carbon-enhanced metal-poor stars at high resolution
(Norris et al. 1997a, 1997b, 2002,
Bonifacio et al. 1998, Hill et al. 2000, Aoki et al. 2002b)  have revealed 
many  more intriguing  results; 
however,  the specific trend
of increase in carbon-enhanced stars with decreasing metallicity still
remains unexplained. Also, the production mechanisms of carbon in these stars 
still remain unknown. There are different types of carbon-enhanced stars;
(i) stars showing carbon enhancement with $s$-process element enhancement,
(ii) carbon enhancement with $r$-process element enhancement
and (iii)   carbon enhancement with normal $n$-capture element abundances.
There is yet another type of very metal-poor stars with strong $s$-process
enhancement but only  slightly carbon-enhanced ([C/Fe] = +0.2;  Hill et 
al.  2002).  Certainly a single well defined production mechanism  is 
unlikely to lead to such a   diversity in  abundances.  To shed light on 
the production mechanisms  of carbon-excess resulting in    different types 
of carbon-enhanced stars and to understand the nucleosynthesis of $s$-process,
and $r$-process  elements at low metallicity  it is desirable to conduct
analysis of as many different types of C-enhanced stars as possible.

Christlieb et al. (2001) reported a sample of 403 Faint High Latitude Carbon
(FHLC) stars identified by means of line indices - i.e. ratios of the mean 
photographic densities in the carbon molecular absorption features and the
continuum band passes, which were the basis for the Hamburg catalog of high
Galactic latitude carbon stars. The identification was primarily based on
the presence of strong C$_{2}$ and CN molecular bands shortward of 5200 \AA\,;
it  did not consider CH bands.
At high galactic latitudes, although the surface density of FHLC stars is 
low, different kinds of carbon stars are known to populate the region
(Green et al. 1994). One kind is the normal asymptotic giant-branch (AGB)
stars, carbon-enriched by dredge-up during post-main-sequence phase which
are found among the N-type carbon stars. Another kind is the  FHLC stars
showing significant proper motions and having luminosities of main-
sequence dwarf called dwarf carbon stars (dCs). A third kind of FHLC stars
is the so-called CH-giant stars, similar to the metal-poor carbon stars
found in Globular clusters and some in dwarf spheroidal (d Sph) galaxies
(Harding 1962). Among these, at high galactic latitudes warm carbon stars
possibly some C-R stars are also likely to be present. The sample of stars
offerred by Christlieb et al. (2001) being high latitude objects, with
smaller initial mass and possible lower metallicity is likely to contain a
mixture of these objects.  
Different kinds of objects have different astrophysical
implications and hence it is important to distinguish them from one
another, although in certain cases it is not easy to do so. For example,
dCs are difficult to distinguish from C-giants as they exhibit remarkable
similarity in their spectra with those of C-giants. They are however
distinguishable through their relatively high proper motion and 
apparently anomalous JHK infrared colours (Green et al. 1992).

Interpretation of chemical compositions  of the intermediate-mass  
stars formed from the interstellar matter is not straight forward  as 
the interstellar matter is already affected by the ejecta of many generations 
of more massive stars. In comparison, the halo red giant stars offer more 
direct information on the role of intermediate-mass stars of the halo. 
Thus,  existance of CH stellar component has  important astrophysical 
implications for  Galactic chemical evolution. The processes responsible 
for carbon excess in these stars to a large extent are responsible for the 
origin and evolution of carbon, nitrogen and heavy elements in the early 
Galaxy.  Furthermore, isotopic ratios of $^{12}$C/$^{13}$C in C and 
C-related stars provide useful probes of nucleosynthesis processes and
their location leading to carbon excess in these stars. To determine
these ratios useful candidates are those with strong isotopic carbon bands
in their spectra; CH stars provide an useful set of candidates.

Determination of the chemical compositions as well as carbon isotopic 
ratios $^{12}$C/$^{13}$C would require high resolution spectroscopy. But 
before this,  a target list of CH stars needs to be generated and this can 
be done from spectral analysis of stars using even low resolution spectroscopy. 
Prompted by this we have undertaken to identify the CH as well as  other 
types of stellar objects in the sample of FHLC stars of Christlieb et al. 
 using low resolution spectroscopy. 
These identifications and the  low resolution spectroscopic analysis of the
candidate CH stars is the main theme of this paper.

Observations and data reductions are described in section 2. In section 3 
we  briefly discuss different types of C stars and their spectral 
characteristics.  JHK photometry of the stars is briefly described
in section 4. Description of the program stars spectra   and results 
are drawn in section 5.  Section 6 contains a brief discussion on the 
atmospheres of candidate CH stars. Concluding remarks  are presented in 
section 7. 

\section{Observation and Data Reduction}
The stars listed in Table 1 (51 stars) and Table 2 (40 stars) have been 
observed with 2-m 
Himalayan Chandra Telescope (HCT)  at the Indian Astronomical Observatory
(IAO), Mt. Saraswati, Digpa-ratsa Ri, Hanle during June 2003 - May 2004.
Spectra of a number of carbon stars such as HD 182040, HD 26, HD 5223,
HD 209621, Z PSc, V460 Cyg and RV Sct are also taken for comparison.
A spectrum of C-R star HD 156074  taken from Barnbaum 
et al.'s (1996) atlas  is also used for comparison.
The spectrograph used is the Himalayan Faint Object Spectrograph 
Camera (HFOSC). HFOSC is an optical imager cum a spectrograph for 
conducting low and medium resolution grism spectroscopy 
(http://www.iiap.ernet.in/iao/iao.html).
The grism and the camera combination used for observation
provided a spectral resolution of  $\sim$ 1330( ${\lambda/\delta\lambda}$ ); 
the observed bandpass ran from about 3800 to 6800 \AA\ .
 
Observations of  Th-Ar hollow cathod lamp taken immediately before and 
after the stellar exposures provided the wavelength calibration. The CCD 
data were reduced  using the IRAF software spectroscopic reduction  packages. 
For each object two spectra were taken each of 15 minutes exposures, 
the two spectra were combined to increase the signal-to-noise 
ratio.  2MASS JHK measurements  for the stars in Table 1 are also listed. 
These measurements are available 
on-line at http://irsa.ipac.caltech.edu/ .
  In Table 2, the objects observed on 2nd and 3rd March, 2004 are
acquired using OMR spectrograph at the cassegrain focus of the 2.3 m Vainu 
Bappu Telescope (VBT) at Kavalur. With a 600 lmm$^{-1}$ grating, we get a 
dispersion of 2.6 \AA\, per pixel. 
The spectra of these
objects cover a wavelength range 4000 - 6100 \AA\,,   at a resolution 
of ${\sim}$ 1000.

\section { Types  of C  stars and their spectral characteristics}

Carbon stars  are classified into different spectral types based on their 
characteristic spectral properties. We briefly discuss here  the main 
characteristics  essential for our purpose.  More detail discussion on this 
can be found in literature including Wallerstein (1998) and references 
therein. Among  the carbon stars, the C-N stars have lower temperatures and 
stronger molecular bands than those of C-R stars.  C-N stars exhibit very 
strong depression of light in the violet part of the spectrum.  They are  
used as tracers of an intermediate age population in extragalactic objects. 
The C-R stars as well as CH stars have warmer temperatures and blue/violet 
light is accessible to observation and atmospheric analysis.  C-N stars  
are easily detected in infrared surveys from their characteristic infrared 
colours.
The majority of C-N stars show ratios of $^{12}$C/$^{13}$C more than 30, 
ranging nearly to 100 while in C-R stars this ratio ranges from 4 to 9. 
The strength/weakness  of CH band in C-rich stars provides a measure of 
the degree of hydrogen deficiency in carbon stars. 

The characteristic behaviour of $s$-process elements in C-stars can also be 
used as an useful indicator of spectral type. The s-process element abundances 
are nearly solar in C-R stars (Dominy 1984); whereas most of the carbon and 
carbon related stars show significantly enhanced abundances of the s-process 
elements relative to iron (Lambert et al. 1986, Green and Margon 1994). 

CH stars are  characterised by strong G-band of CH in their spectra. These 
stars are not a homogeneous group of stars. They consist of two populations, 
the most metal-poor ones have a spherical distribution and the ones slightly 
richer in metals are characterised by a flattened ellipsoidal distribution 
(Zinn 1985).  These  stars form a group of warm stars of equivalent spectral 
types G and K giants, but show weak metallic lines. The ratio of the local 
density of CH stars is as high as 30\% of metal-poor giants (Hartwick \& 
Cowley 1985);  and being  the most populous type of halo carbon stars known,
are important objects  for our understanding of galactic chemical evolution,
the evolution of low mass stars and nucleosynthesis in metal poor stars.

Most of the CH stars are known to be high velocity objects. `CH-like' stars, 
where CH are less dominant have low space velocities Yamashita (1975).
At low resolution to make a  distinction between CH and C-R stars is difficult
as many C-R stars also show quite strong CH band. In such cases secondary 
P-branch head near 4342 \AA\, is used as a more  useful indicator.
Another  important  feature is the strength of Ca I at 4226 \AA\, which in 
case of CH stars is weakened by the overlying faint bands of the CH band 
systems. In C-R star this feature is  quite strong. These spectral 
characteristics allow for an identification of CH  and C-R stars even at 
low resolution.  Enhanced lines of s-process elements, weaker Fe group 
elements as well as various strengths of C$_{2}$ bands are some  other
distingushing spectral features of CH stars. However, at low dispersion the 
narrow lines are difficult to estimate and essentially do not provide with 
a strong  clue to  distinguish  C-R stars from  CH stars. Although CH and 
C-R stars have similar range of temperatures the distribution of CH stars 
place most of them in the Galactic halo, their large radial velocities ,
typically $\sim$ 200km s$^{-1}$ are indicative of their being halo objects
(McClure 1983, 1984).
 
The objects observed from Hanle are classified considering these spectral 
characteristics.  In the following  we  discuss the medium resolution 
spectra of the objects listed in Table 1 with their photometric data.

\begin{figure*}
\epsfxsize=14truecm
\epsffile{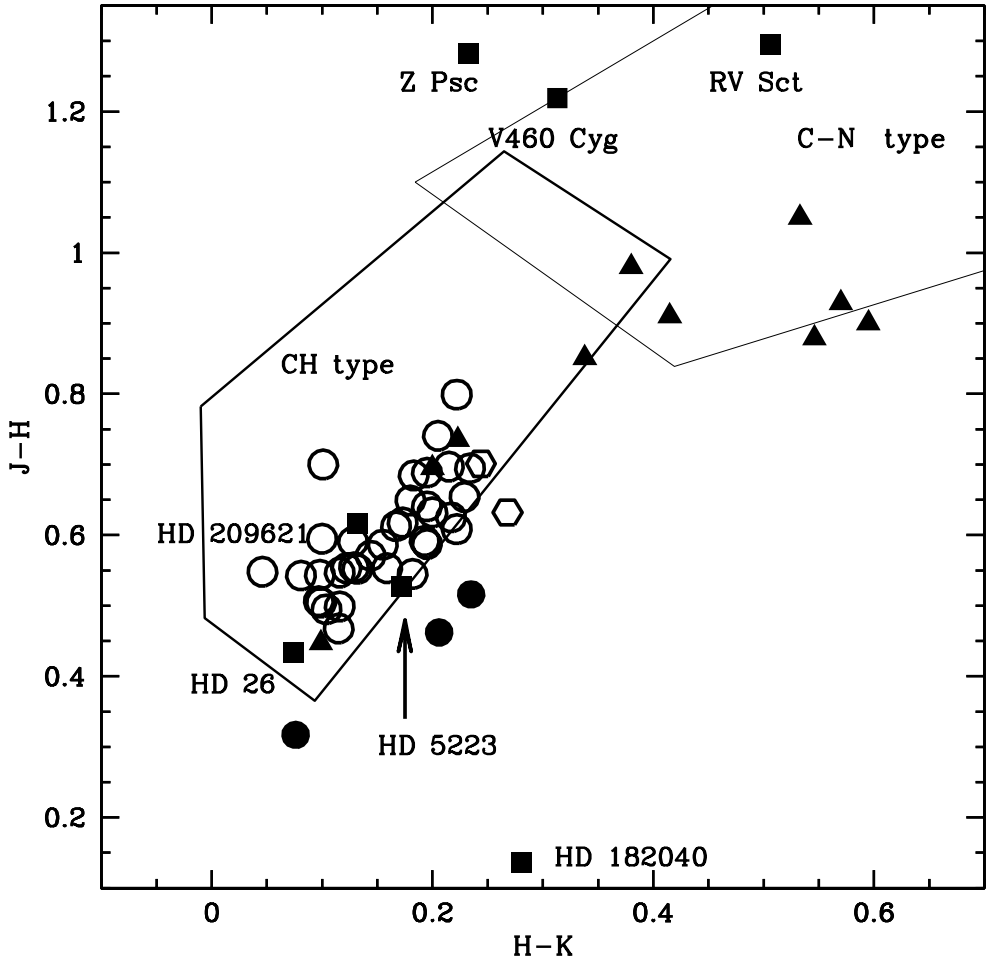}
\caption{
A two colour J-H versus H-K diagram  of the  stars listed in Table 1. 
The candidate CH stars are represented by open circles except the three 
outliers represented by solid circles. C-N stars are represented 
by solid triangles  and C-R stars by open hexagon. The two boxes 
superimposed in the figure illustrate
the loci of separate carbon-star types and are taken from Totten et al.
(2000). The location of the comparison stars are labeled and maked with 
solid squares.}
\label{Figure 1}
\end{figure*}

{\footnotesize
\begin{table*}
{\bf Table 1: HE  stars with prominent C$_{2}$ molecular bands }\\
\begin{tabular}{cccccccccccccc}
             &        &      &          &    &   &             &          &      &      &       &        &       &\\
\hline
Star No.   & RA(2000)$^{a}$ & DEC(2000)$^{a}$& $l$ & $b$ & B$_{J}^{a}$& V$^{a}$ & B-V$^{a}$  &  U-B$^{a}$ &J & H & K &    &Dt of Obs\\
\hline
            &           &          &      &       &     &     &     &     &      &       &   &     &   \\
HE 0002+0053& 00 05 25.0& +01 10 04& 99.71& -59.61& 14.5& 13.3& 1.72& 1.25& 11.018&10.386&10.118&         & 06.11.04\\
HE 0017+0055& 00 20 21.6& +01 12 07&106.90& -60.70& 12.6&     &     &     &  9.309& 8.693& 8.498&         & 15.11.03 \\
HE 0038-0024& 00 40 48.2& -00 08 05&117.09& -62.89& 15.4& 14.4& 1.86& 1.67& 12.433&11.768&11.573&         & 06.11.04\\
HE 0043-2433& 00 45 43.9& -24 16 48& 98.33& -86.88& 13.8& 13.1& 1.04& 1.00 &11.064&10.493&10.365&        & 07.11.04\\
HE 0110-0406& 01 12 37.1& -03 50 30&136.11& -66.17& 13.4&     &     &     & 10.523& 9.988& 9.866&        & 17.9.03\\
HE 0111-1346& 01 13 46.5& -13 30 49&145.01& -75.42& 13.3&     &     &     & 10.684&10.155&10.039&         & 07.11.04\\
HE 0151-0341& 01 53 43.3& -03 27 14&157.78& -62.04& 14.6& 13.4& 1.27& 0.87& 11.847&11.364&11.248&         & 07.11.04\\
HE 0207-0211& 02 10 12.0& -01 57 39&163.12& -58.55& 15.5& 14.0& 2.16& 2.13& 11.505&10.605&10.010&       & 07.11.04\\
HE 0308-1612& 03 10 27.1& -16 00 41&201.12& -55.96& 12.5&     &     &     & 10.027& 9.475& 9.331&       & 17.9.03 \\
HE 0310+0059& 03 12 56.9& +01 11 10&178.95& -45.73& 12.6&     &     &     &  9.871& 9.296& 9.196&         & 17.9.03\\
HE 0314-0143& 03 17 22.2& -01 32 37&182.98& -46.69& 12.7&     &     &     &  8.993& 8.222& 8.000&          & 17.9.03 \\
HE 0319-0215& 03 21 46.3& -02 04 34&184.58& -46.17& 14.6& 13.6& 1.43& 1.01& 11.785&11.218&11.063&           & 16.9.03 \\
HE 0322-1504& 03 24 40.1& -14 54 24&201.90& -52.39& 15.0& 13.8& 1.63& 1.24& 12.105&11.533&11.340&         & 06.11.04\\
HE 0429+0232& 04 31 53.7& +02 39 01&192.72& -29.17& 14.2& 13.3& 1.35& 1.08 &11.088&10.520&10.325&         & 07.11.04\\
HE 0457-1805& 04 59 43.6& -18 01 11&217.85& -32.51& 12.1& 11.2& 1.25& 1.20 & 8.937& 8.421& 8.186&         & 07.11.04\\
HE 0507-1653& 05 09 16.5& -16 50 05&217.54& -29.96& 15.6& 12.4& 1.06& 0.68& 10.883&10.430&10.315&         & 06.11.04\\
HE 0518-2322& 05 20 35.5& -23 19 14&225.62& -29.74& 13.7&     &     &      &11.151&10.672&10.568&         & 15.11.03\\
HE 0915-0327& 09 18 08.2& -03 39 57&235.26& +30.09& 14.5& 12.9& 2.29& 2.12&  9.968& 8.989& 8.609&          & 10.4.04\\
HE 0932-0341& 09 35 10.2& -03 54 33&238.38& +33.41& 14.8& 13.9& 1.23& 1.02 &12.295&11.807&11.708&         & 06.11.04\\
HE 1008-0636& 10 10 37.0& -06 51 13&248.12& +38.35& 14.5& 12.9& 2.28& 2.11&  9.952& 9.073& 8.527&         & 29.3.04\\
HE 1027-2501& 10 29 29.5& -25 17 16&266.68& +27.42& 13.9& 12.7& 1.73& 1.51&       &      &      &         & 30.3.04\\
HE 1056-1855& 10 59 12.2& -19 11 08&269.48& +36.29& 13.6&     &     &      &10.784&10.249&10.090&         & 20.12.04\\
HE 1104-0957& 11 07 19.4& -10 13 16&265.35& +44.92& 14.7&     &     &     &  8.262& 7.561& 7.317&         & 20.12.04\\
HE 1107-2105& 11 09 59.6& -21 22 01&273.53& +35.65& 14.3& 12.1& 3.11& 2.44&  8.279& 7.229& 6.696&         & 30.3.04\\
HE 1125-1357& 11 27 43.0& -14 13 32&274.20& +43.93& 15.2& 14.1& 1.41& 1.40& 11.730&11.057&10.842&         & 12.4.04\\
HE 1145-0002& 11 47 59.8& -00 19 19&271.30& +58.60& 13.5& 13.6& 1.48& 1.49& 10.911&10.240&10.006&         & 11.4.04\\
HE 1204-0600& 12 07 11.6& -06 17 06&283.56& +54.91& 14.9& 14.0& 1.36& 1.45& 11.517&10.898&10.703&       & 11.4.04\\
HE 1211-0435& 12 14 12.0& -04 52 26&285.83& +56.76& 15.0& 14.2& 1.08& 0.90& 12.492&11.962&11.916&       & 12.4.04\\
HE 1228-0402& 12 30 50.6& -04 18 59&293.16& +58.16& 16.3& 15.1& 1.68& 1.92& 12.805&12.070&11.847&        & 11.4.04\\
HE 1254-1130& 12 56 57.0& -11 46 19&305.08& +51.08& 16.1& 14.5& 2.13& 2.37& 10.731& 9.821& 9.406&       & 30.3.04\\
HE 1259-2601& 13 01 52.4& -26 17 16&305.84& +36.52& 13.9& 12.8& 1.77& 1.56&       &      &      &       & 03.3.04\\
HE 1304-2046& 13 06 50.1& -21 02 10&307.75& +41.69& 15.2& 14.3& 1.32& 1.36& 11.978&11.386&11.219&      & 30.3.04\\
HE 1305+0132& 13 08 17.8& +01 16 49&312.52& +63.84& 13.8& 12.8& 1.35& 1.25& 10.621& 9.994& 9.814&       & 28.3.04\\
HE 1418+0150& 14 21 01.2& +01 37 18&346.80& +56.66& 14.2&     &     &     &  9.988& 9.356& 9.127&       & 10.4.04 \\
HE 1425-2052& 14 28 39.5& -21 06 05&331.40& +36.64& 13.6& 12.7& 1.27& 1.29& 10.043& 9.446& 9.273&       & 28.3.04\\
HE 1429-0551& 14 32 31.3& -06 05 00&343.02& +48.76& 13.5&     &     &     & 10.734&10.272&10.066&       &05.9.03\\
HE 1446-0112& 14 49 02.2& -01 25 24&352.42& +49.80& 14.5& 13.5& 1.38& 1.39& 10.983&10.379&10.162&       &06.9.03\\
HE 1501-1500& 15 04 26.3& -15 12 00&344.28& +36.78& 16.5& 15.3& 1.65& 1.61& 12.725&12.030&11.830&       & 10.4.04\\
HE 1523-1155& 15 26 41.0& -12 05 43&351.87& +35.63& 14.2& 13.4& 1.14& 0.70& 11.372&10.846&10.748&       & 29.3.04\\
HE 1524-0210& 15 26 56.9& -02 20 45& 0.98 & +42.35& 14.4& 13.3& 1.53& 1.25& 11.740&11.079&10.896&       & 06.9.03\\
HE 1528-0409& 15 30 54.3& -04 19 40&359.87& +40.30& 15.8& 15.0& 1.10& 0.78& 12.945&12.455&12.358&       & 29.3.04\\
HE 2144-1832& 21 46 54.7& -18 18 15& 34.65& -46.78& 12.6&     &     &     &  8.768& 8.180& 7.958&        & 16.9.03\\
HE 2145-1715& 21 48 44.5& -17 01 03& 36.63& -46.73& 14.2& 13.2& 1.39& 1.18& 11.032&10.356&10.255&        & 17.9.03\\
HE 2207-0930& 22 09 57.5& -09 16 06& 50.27& -47.96& 14.4& 13.1& 1.82& 1.40& 10.527& 9.812& 9.607&        & 16.9.03\\
HE 2207-1746& 22 10 37.5& -17 31 38& 38.87& -51.77& 11.8&     &     &     &  9.115& 8.579& 8.450&        & 06.9.03\\
HE 2218+0127& 22 21 26.1& +01 42 20& 65.46& -43.80& 14.6& 14.0& 0.80& 0.31& 11.826&11.509&11.433&         & 16.9.03\\
HE 2221-0453& 22 24 25.7& -04 38 02& 59.04& -48.38& 14.7& 13.7& 1.36& 1.11& 11.524&10.997&10.815&         & 17.9.03\\
HE 2239-0610& 22 41 53.1& -05 54 22& 61.61& -52.61& 14.1& 13.1& 1.34& 1.59 &13.830&13.296&13.164&         & 07.11.04\\
HE 2319-1534& 23 22 11.1& -15 18 16& 58.09& -66.14& 15.3& 13.8& 2.09& 2.16& 10.866& 9.937& 9.367&         & 17.9.03\\
HE 2331-1329& 23 33 44.5& -13 12 34& 66.55& -67.12& 16.2& 14.5& 2.29& 2.19 &11.841&10.990&10.652&         & 06.11.04\\
HE 2339-0837& 23 41 59.9& -08 21 19& 78.51& -65.05& 14.9& 14.0& 1.32& 0.62 &12.632&12.107&12.026&         & 06.11.04\\
\hline

\end{tabular}

$^{a}$ From Christlieb et al. (2001)\\
\end{table*}
}

\begin{table*}
{\footnotesize
\small
{\bf Table 2: HE stars without prominent $C_{2}$ bands  }\\
\begin{tabular}{ccccccccccc}
\hline
Star No.    &  RA(2000)$^{a}$ & DEC(2000)$^{a}$& $l$ & $b$ &  B$^{a}$ & V$^{a}$ & B-V$^{a}$ & U-B$^{a}$ &Bands          & Dt of Obs\\
            &                 &                &     &     &          & 
&           &           & noticed       &          \\
\hline
HE 0201-0327& 02 03 49.0& -03 13 05&161.94& -60.49& 14.1& 13.4& 1.02& 0.95 & CH, CN     & 07.11.04  \\
HE 0333-1819& 03 35 18.8& -18 09 54&208.37& -51.32& 12.6&     &     &     & CH, CN         & 16.9.03 \\
HE 0359-0141& 04 02 21.2& -01 33 05&192.03& -37.64& 14.5& 13.4& 1.26& 1.08 &CH, CN         & 15.11.03\\
HE 0408-1733& 04 11 06.0& -17 25 40&211.87& -43.11& 13.1& 12.2& 1.28& 1.26& CH, CN          & 17.9.03\\
HE 0417-0513& 04 19 46.8& -05 06 17&198.66& -35.82& 14.6& 13.7& 1.31& 1.21 &CH, CN         & 15.11.03\\
HE 0419+0124& 04 21 40.4& +01 31 46&192.17& -31.92& 15.7& 13.0& 1.44& 1.37 & CH, CN     & 07.11.04  \\
HE 0443-1847& 04 46 10.9& -18 41 40&217.23& -35.75& 13.1& 12.9& 1.27& 1.21& CH, CN         & 16.9.03\\
HE 0458-1754& 05 00 34.5& -17 50 21&217.73& -32.26& 13.5& 12.7& 1.18& 1.09 &CH, CN         & 02.3.04\\
HE 0508-1604& 05 10 47.0& -16 00 40&216.82& -29.31& 12.8& 12.1& 1.04& 1.15 & CH, CN,    & 20.12.04  \\
HE 0518-1751& 05 20 28.4& -17 48 43&219.71& -27.84& 13.5& 12.8& 1.05& 1.22 & CH, CN    & 07.11.04  \\
HE 0519-2053& 05 21 54.4& -20 50 36&223.06& -28.62& 13.6& 13.7& 1.18& 1.14 &CH, CN         & 15.11.03\\
HE 0536-4257& 05 37 40.4& -42 55 39&248.71& -31.11& 13.8& 12.7& 1.44& 1.41 &               & 03.3.04\\
HE 0541-5327& 05 42 14.3& -53 26 31&261.05& -31.59& 13.6&     &     &      &               & 03.3.04\\
HE 0549-4354& 05 50 34.3& -43 53 24&250.28& -28.98& 13.7& 12.8& 1.31& 1.18 & CH            & 03.3.04\\
HE 0900-0038& 09 02 50.5& -00 50 20&230.15& +28.43& 14.2& 13.3& 1.27& 1.19 & CH, CN        & 29.3.04\\
HE 0916-0037& 09 18 47.6& -00 50 35&232.63& +31.79& 13.7& 12.8& 1.24& 1.02 & CH            & 03.3.04\\
HE 0918+0136& 09 21 26.1& +01 23 28&230.81& +33.55& 14.0& 13.1& 1.30& 1.21 & CH            & 03.3.04\\
HE 0919+0200& 09 22 13.0& +01 47 56&230.52& +33.93& 13.5& 12.6& 1.31& 1.20 & CH            & 03.3.04\\
HE 0930-0018& 09 33 24.7& -00 31 46&234.74& +35.01& 14.2& 14.7& 1.43& 1.45 &CH             & 02.3.04\\
HE 0935-0145& 09 37 59.0& -01 58 36&236.99& +35.12& 13.8& 12.9& 1.16& 1.07 &CH             & 02.3.04 \\
HE 0939-0725& 09 42 11.9& -07 39 06&243.19& +32.50& 14.0& 13.1& 1.20& 1.13 & CH, CN,   & 20.12.04\\
HE 1042-2659& 10 44 24.2& -27 15 30&271.05& +27.64& 14.7& 12.6&     &      & CH            & 03.3.04\\
HE 1117-2304& 11 19 42.8& -23 21 07&277.08& +34.87& 13.3&     &     &      & CH, CN        & 11.4.04\\
HE 1119-3229& 11 22 21.9& -32 46 19&282.08& +26.47& 14.0& 13.1& 1.18& 1.25 & CH            & 03.3.04\\
HE 1227-3103& 12 30 34.5& -31 19 54&297.72& +31.33& 14.3& 13.3& 1.39& 1.54 &               & 02.3.04\\
HE 1304-3020& 13 07 24.2& -30 36 36&306.99& +32.14& 13.5& 12.7& 1.17& 1.06 &CH             & 02.3.04\\
HE 1356-2752& 13 59 25.0& -28 06 59&320.71& +32.40& 13.3&     &     &      & CH            & 03.3.04\\
HE 1455-1413& 14 57 51.6& -14 25 10&343.27& +38.36& 13.1&     &     &      & CH            & 03.3.04\\
HE 1500-1101& 15 03 40.9& -11 13 09&347.25& +40.01& 13.8& 12.9& 1.28& 1.24 & CH            & 29.3.04\\
HE 1514-0207& 15 16 38.9& -02 18 33&358.67& +44.29& 13.6&     &     &     & CH, CN         & 05.9.03\\
HE 1521-0522& 15 24 12.2& -05 32 52&357.20& +40.70& 14.7& 13.8& 1.24& 1.11 & CH, CN        & 11.4.04\\
HE 1527-0412& 15 29 42.3& -04 22 22&369.56& +40.49& 13.8& 12.9& 1.21& 1.19& CH, CN         & 05.9.03\\
HE 2115-0522& 21 18 11.8& -05 10 07& 46.39& -34.80& 17.4& 14.3& 1.22& 1.15 & CH, CN        & 07.11.04  \\
HE 2121-0313& 21 23 46.2& -03 00 51& 49.51& -34.90& 14.9& 13.9& 1.35& 1.47& CH, CN         & 05.9.03\\
HE 2124-0408& 21 27 06.8& -03 55 22& 49.09& -36.09& 14.8& 13.9& 1.26& 1.15& CH, CN         & 17.9.03\\
HE 2138-1616& 21 41 16.6& -16 02 40& 36.95& -44.70& 14.7& 13.9& 1.01& 0.91& CH, CN         & 16.9.03\\
HE 2141-1441& 21 44 25.7& -14 27 33& 39.43& -44.77& 14.3& 13.5& 1.13& 1.03& CH, CN         & 16.9.03\\
HE 2145-0141& 21 47 48.3& -01 27 50& 55.23& -39.10& 13.4& 12.6& 1.10& 1.02& CH, CN         & 16.9.03\\
HE 2224-0330& 22 26 47.9& -03 14 58& 61.23& -48.01& 14.3& 13.5& 1.08& 0.94& CH, CN         & 16.9.03\\
HE 2352-1906& 23 54 49.0& -18 49 31& 62.50& -74.57& 12.9&     &     &      &CH, CN         & 16.9.03\\
\hline

\end{tabular}

$^{a}$ From Christlieb et al. (2001)\\
}
\end{table*}

\section {JHK photometry}

Infrared colours made from JHK photometry provide a supplementary diagnostics
for stellar classification. Figure 1 is a two colour JHK diagram where
J-H versus H-K colours of  HE stars listed in Table 1 are plotted.
The HE stars 2MASS JHK measurements are available on-line at
http://iras.ipac.caltech.edu/.

The two boxes superimposed in the figure representing the location of CH 
stars (thick line solid box) and the C-N stars (thin line solid box)
illustrate the loci of the separate carbon-star types and are taken from
Totten et al. (2000). In this figure,  the CH stars classified by us
(following our  discussions in the subsequent sections),  plotted with
open circles fall well within the CH box, except the three outliers
HE 1429-0551, HE 2218+0127 and HE 0457-1805. These three  stars are 
represented by solid
circles. The spectral characteristics of these  stars led us to classify
them as CH stars. Their spectra do not show any peculiarities from  
which  their location in the J-H, H-K plane seems obvious. A difference 
between the spectra of the first two stars  lies in molecular C$_{2}$ bands 
in the spectral region 5700 - 6800 \AA\, . In this region HE 1429-0551 
does not show molecular C$_{2}$ bands (or could be marginally detected) 
whereas HE 2218+0127 shows molecular C$_{2}$ bands as strongly (or marginally 
stronger) as they are seen in CH star HD 5223. Ba II feature at 6496 \AA\, 
is weak in HE 1429-0551. In HE 2218+0127, this feature  appears to be of 
equal depth to its counterpart in HD 5223. HE 2218+0127 seems 
to be the warmest  among the candidate  CH stars (Table 3).
HE 0457-1805, another CH star   outside the CH box resembling HD 26, 
a known CH star,  shows stronger CN molecular band
around 4215 \AA\, and slightly stronger features due to Ba II at 6496 \AA\,
and Na I D. H$_{\alpha}$ feature is marginally weaker but G-band of CH
appears almost of equal strength.  There are ten stars in the present 
sample which show spectral characteristics
of C-N stars, they are represented by solid triangles. Four of them fall
well within the C-N box, three of them just outside the C-N box
 and the rest two fall within the CH box. Stars 
HE 2319-1534 and HE 1008-0636 at the redder edge of the C-N box show
H$_{\alpha}$ and H$_{\beta}$ in emission whereas HE 2331-1329, HE 0915-0327 
and HE 1254-1130
with lower H-K values do not show H$_{\alpha}$ and H$_{\beta}$ features
in their spectra. HE 1501-1500,  HE 1228-0402 and HE 1107-2005 (inside 
the CH box) do
not have flux below 4500 \AA\,. CN molecular bands are weaker in HE 1228-0402
than their counterparts in other C-N stars. This is not the case with
HE 1501-1500.  H$_{\alpha}$ and H$_{\beta}$ features are not detectable in
these two stars. At present it remains to be understood why these two 
stars occupy a location among the CH stars in the J-H, H-K plane.

\section{ Results}

\subsection{ Spectral characteristics of the program stars}

The spectra are examined in terms of the following  spectral characteristics.\\
1. The strength (band depth) of CH band around  4300 \AA\,.\\
2. Prominance of Secondary P-branch head near 4342 \AA\,.\\
3. Strength/weakness  of Ca I feature at 4226 \AA\,.\\
4. Isotopic band depths  of C$_{2}$ and CN,  in particular the Swan bands 
 of $^{12}$C$^{13}$C and $^{13}$C$^{13}$C near 4700 \AA\,.\\
5. Strength of other C$_{2}$ bands in the 6000 -6200 \AA\, region. \\  
6. $^{13}$CN band near 6360 \AA\, and other CN bands across the wavelength
range.\\
7. Strength of s-process element  such as Ba II features at 4554 \AA\, and 
 6496 \AA\,.

To establish 
the   membership of a star in a particular  group we have conducted a 
differential analysis of the program stars spectra with the spectra
of carbon stars  available in the low resolution spectral atlas of carbon 
stars of  Barnbaum et al. (1996).  We have also acquired spectra for
some of the objects from this Atlas and used them for comparison of
spectra at the same resolution.

\subsection{Candidate CH  stars: Description of the spectra} 

At low resolution  the spectra of C-R and CH stars look very similar  and 
this makes  distinction between them  a difficult  task. The differences  
are made apparent by making a comparison between spectra of  known C-R  
and CH stars.  Application of this comparison to the program stars helped 
in an easy identification of their spectral class. In figures 2 and  3  we
show a comparison of the spectra of a pair of C-R stars  HD 156074 and HD 76846 
and a pair of CH stars HD209621 and HD 5223.   
Although we have considered here four stars,
 the comparison is generally true for any C-R and CH stars.\\

{\it A comparison of known   C-R and CH stars spectra }\\

(i) Wavelength region 4000 - 5400 \AA\, (Figure 2):\\
G-band of CH is strong in  all the spectra, almost of equal strength. However, 
the secondary P-branch head around 4343 \AA\, is distinctly seen in the CH
stars spectra. In C-R stars spectra this feature is   merged with contributions
from  molecular bands. 

 In C-R stars the Ca I  at 4226 \AA\, line depth  is almost equal 
to the CN band depth at 4215 \AA\, whereas in CH stars spectra this line
is marginally noticed. CN band around 4215 \AA\, is much deeper in C-R
stars than in the CH stars.

 Narrow atomic lines are blended with contributions from molecular bands
and hence their real strength could not be estimated at this resolution.
In the above wavelength range 
 H$_{\beta}$ and Ba II at 4554 \AA\, are the  two features
clearly noticeable in the CH stars. In C-R star this region is 
a complex combination of atomic and molecular lines. 
There is no obvious distinction in the isotopic bands around 4700 \AA\,
in C-R and CH stars.
C$_{2}$ molecular bands around 5165 \AA\, and 5635 \AA\, are two  
prominent features in this region.  

(ii) Wavelength region 5400 -6800 \AA\, (Figure 3)\\
C$_{2}$ molecular bands around 5635 \AA\,  is the most    
prominent feature in this region.  
This region too is a complex mixture of atomic and molecular lines.
 A blended feature of Na I D$_{1}$ and Na I D$_{2}$ in C-R stars is sharper 
with two distinct dips. In CH stars this feature is shallower and the 
individual contribututions of  Na I D$_{1}$ and Na I D$_{2}$ are not 
distinguishable. 
H$_{\alpha}$ feature appears as a  distinct 
feature in CH stars; in C-R stars this feature seems to be contaminated
by molecular contributions. Ba II feature at 6496 \AA\, is also 
blended with contributions from CN bands around 6500 \AA\,; in CH stars this
blending is not so severe. 
CN molecular bands, although present are in general weaker in CH stars 
than in C-R stars.
 
The main features of the above comparison are   used  to identify the
spectral type ( CH or  C-R ) of the program stars. A small number of  C-N stars
were  easily identified from their distinct spectral properties. In figure
4  we present the spectra of the comparison stars in the wavelength
region 4000 -6800 \AA\,. In figure 5 we show one example of HE stars 
corresponding to each comparison star's spectrum in figure 4, in
the sequence top to bottom.
In the following  we present the spectral description of the individual star.\\

{\bf HE 2145-1715, 
 HE 0518-2322, HE 0457-1805, HE 0043-2433, HE 1056-1855}\\
The spectra of these objects closely resemble the spectrum of HD 26,
 a known CH star. 
 CH bands around ${\lambda}$4300 are of almost equal strength in the   
spectra of these stars.  Ca I 4226 \AA\, line is very weak, 4271 Fe I line 
is barely  
detectable. Strength of G-band of CH, prominent secondary P-branch head 
around 4342 \AA\, and a weak Ca I feature at 4226 \AA\, show that these stars 
could be  CH stars.

C$_{2}$ molecular bands around 4730 \AA\,, 5165 \AA\, and 5635 \AA\, are much
deeper  in HE 2145-1715 than their counterparts in HD 26. H$_{\beta}$ 
features are 
of equal strength. Ba II line around 4545 \AA\, is marginally weaker in the
spectrum  of HE 2145-1715 whereas Ba II feature at 6496 \AA\, and H$_{\alpha}$
are of 
equal strength.  The effective temperature of HD 26 is ${\sim}$ 4880 K,  
and [Fe/H]=-0.5 (Aoki \& Tsuji 1997).  A marginally weaker NaI D feature than 
in HD 26 spectrum  and the deeper C$_{2}$ bands in HE 2145-1715  perhaps is 
an indication of slightly lower metallicity and lower temperature  for 
HE 2145-1715 than  HD 26. This statement however can be asscertained only 
from high resolution spectral analysis.\\
 In HE 0518-2322, CN molecular band depth matches well with that of HD 26.
Na I D appears weakly in emission, Ba II at 6496 \AA\, and H$_{\alpha}$
features are marginally stronger. H$_{\alpha}$ feature has an weak emission
at the absorption core. HE 0043-2433 has a stronger  CN band around 4215 \AA\,
but H$_{\alpha}$, H$_{\beta}$ and Ba II at 6496 \AA\, appear with almost
similar strength to those in  HD 26.  Na I D feature appears weakly in
absorption in this star. In HE 0457-1805,  Na I D is stronger than in HD 26
but H$_{\alpha}$, H$_{\beta}$ and Ba II at 6496 \AA\, appear with almost
similar strength. In HE 1056-1855,  H$_{\alpha}$ and  H$_{\beta}$ are 
marginally weaker but Ba II at 6496 \AA\, appear with almost equal strength
as in HD 26.\\

{\bf HE 0310+0059,  HE2239-0610, HE 0932-0341, HE 0429+0232 }\\
 These  four stars spectra resemble the spectrum of HD 26  to a large
extent.  G-band of CH  around 4300 \AA\, is of similar strength to that 
in HD 26 but the secondary P-branch head around 4342 \AA\, is not seen
prominently as it is seen in CH stars. Further, in contrast to HD 26, 
 these stars spectra  
 exhibit   strong Ca I feature at 4226 \AA\, in their spectra. These stars 
do not seem to be  potential candidate  CH stars.
In HE 0310+0059,  lines appear much  sharper than in HD 26 and especially
Na I D feature is seen as a much  stronger feature  in absorption. 
In HE 0429+0232,
this feature is marginally weaker than in HD 26.
In    HE 2239-0610 and HE 0932-0341
Na I D  features appear in weak emission. CN bands are stronger in HE 0310+0059
but C$_{2}$ bands are of similar strength. H$_{\alpha}$, H$_{\beta}$,
and Ba II feature at 6496 \AA\, appear  in these stars almost with equal 
strength as in HD 26. \\

{\bf HE 0110-0406, HE 0308-1612, HE 0314-0143, HE 1125-1357, 
HE 1211-0435, HE 1225-2052, HE 1446-0112, HE 1524-0210, HE 1528-0409, 
HE 2144-1832,  HE 2207-1746 }\\
The spectra of these stars closely resemble the spectrum of HD 209621 
except for some marginal differences in the molecular band depths. 
 The star HD 209621 
is a known CH giant with effective temperature $\sim$ 4700 K and metallicity
-0.9 (Wallerstein 1969, Aoki \& Tsuji 1997). 

Except for HE 1446-0112 and HE 1524-0210 the CN band depth around 
${\lambda}$4215 are weaker in the program stars spectra than in the spectrum 
of HD 209621.  Ca I at 4226 \AA\, is not detectable in the spectra of 
HE 1446-0112,  HE 1127-1357, and HE 1211-0435, but appears  weakly in the 
rest of the  stars  spectra.
In the first three stars although Ca I feature at 4226 \AA\, is seen
with its depth almost half the depth of CN band around 4215 \AA\,
it should be noted that in these three stars CN band itself is much
weaker than its counterpart in HD 209621 and in C-R stars.
CH band at ${\lambda}$4300 in the spectra of the program stars  
are  equal or stronger than  in the spectrum of HD 209621 
 except for  HE 2207-1746, HE 0308-1612 and HE 0110-0406 where
this features are slightly weaker. In these stars CN band around 4215 \AA\,
is also weak, much weaker than in C-R stars.
Secondary P-branch head around 4342 \AA\, is  seen  prominently 
in all the cases.  We assign the membership  of these stars to the CH group.
Molecular band heads of C$_{2}$ around ${\lambda}$4700 is of equal strength 
in HE 1446-0112, HE 1524-0210 and HE 1127-1357; in the rest of the stars 
spectra this
band is slightly weaker than in HD 209621. C$_{2}$ band depth 
around ${\lambda}$5165 and ${\lambda}$5635 are almost of equal strength
except for stars HE 2207-1746, HE 1211-0435,  HE 0308-1612, and HE 0110-0406.
Ba II feature at 4554 \AA\, is detectable and of similar strength; 
 however H$_{\beta}$ feature is weaker in HE 1524-0210 and
HE 1528-0409. Except in  HE 1446-0112 and HE 1528-0409 where  
NaI D feature appears  slightly weaker, in  the  rest of the  stars
spectra  this feature is of similar strength with that of NaI D feature 
in HD 209621.
 Ba II feature at 6496 \AA\, which is distinctly seen in HD 209621 
appears blended with CN molecular band in HE 1446-0112. In HE 1125-1357,
HE 1528-0409 and HE 1211-0435 this feature appears slightly weaker than
in HD 209621 and in the rest they  seem to be of equal strength. H${\alpha}$
profile is of equal strength in all the stars except in  HE 0314-0143
where this feature is slightly  weaker.
In figure 6, we show  as an example a comparison of spectra of three objects
in the wavelength region  4125 - 5400 \AA\,  with the spectrum of HD 209621. \\

{\bf HE 1429-0551, HE 1523-1155, HE 2218+0127, HE 2221-0453,  
HE 1204-0600, HE 1418+0150, HE 2207-0930, HE 1145-0002, HE 0111-1346, 
HE 0151-0341, HE 0507-1653, HE 0038-0024, HE 0322-1504, HE 2339-0837}\\
With marginal differences in the molecular band depths these stars spectra 
closely resemble the spectrum of HD 5223,  a well known CH giant 
with effective temperature ${\sim}$ 4500 K, and metallicity -1.3 
(Aoki \& Tsuji 1997). 

The CN band depth around ${\lambda}$4215  in the HE stars
spectra are very similar to the CN band depth  in the spectrum of 
HD 5223 except in HE 1145-0002 where this feature is weaker and does not
show a sharp clear band head.  G-band of CH  around ${\lambda}$4300
in the spectra of the program stars  resemble greatly to their counterparts 
in HD 5223.  Ca I at 4226 \AA\, is  seen in the spectra of 
HE 2218+0127,  HE 1204-0600 and HE 2207-0930 but not  as prominently
as they are seen in C-R stars. Moreover, the line depth of this feature
is quite shallow compared to the CN molecular band depth around 4215 \AA\,.
We note, in C-R stars these two features appear almost with equal depth
and CN band depth is deeper in C-R stars  than in CH stars. 

 The Ca I feature is seen marginally also  in the rest of the  stars  spectra.
Fe I at 4271.6 \AA\, although  weak  could be marginally 
detected in all the spectra. Prominance of secondary P-branch head near
4342 \AA\,, strong G-band of CH and weak or marginally detectable Ca I
feature at 4226 \AA\, allow these stars to be placed in CH group.
The dominance of  CH is shown not only by the marked band depths, but also
by the weakness of Ca I at  4226 \AA\, and distortion of metallic lines
between 4200 and 4300 \AA\,. 
In figure 7, we show  a comparison of  three   spectra in the 
wavelength region  4000 - 5400 \AA\,  with the spectrum of HD 5223.  

Isotopic bands of Swan system around ${\lambda}$4700 appear to be
of equal strength in HE 1204-0600 and   HE 2218+0127 with their counterpart 
in HD 5223. These bands are slighly deeper in HE 2207-0930, HE 1145-0002 
and HE 2221-0453  and marginally swallower in HE 1429-0551 and HE 1523-1155. 
C$_{2}$ bands  around ${\lambda}$5165 and ${\lambda}$5635 greatly resemble  
those  in the spectrum of HD 5223, except for stars HE 2207-0930 and 
HE 1145-0002  where these  bands are slightly  deeper. As in the case of 
HD 5223, Ba II feature at 4554 \AA\, is distinctly seen in the program stars
spectra. However in HE 1429-0551, HE 1523-1155 and HE 2218+0127 this feature 
is marginally weaker, and in the rest the  feature is  of   similar strength. 
H$_{\beta}$ feature  appears in all the spectra with   similar strength  as 
in  HD 5223.  Except in  HE 1429-0551,  HE 1523-1155, and HE 2121-0453,   
NaI D feature appears  slightly stronger  as compared to this 
feature  in HD 5223.  Ba II feature at 6496 \AA\, appears weaker in 
HE 1429-0551, HE 1523-1155 and HE 2218-0127 than in HD 5223,  this 
feature  appears blended with contributions from CN molecular bands in 
HE 1204-0600, HE 2207-0930
and HE 1145-0002.  H${\alpha}$ profile is of equal strength in  HE 1429-0551, 
HE 1523-1155, HE 2218-0127, and HE 2221-0453;  this feature appears slightly 
weaker in HE 1204-0600, HE 2207-0930 and HE 1145-0002 and  blended with 
contributions from molecular bands. The spectra of  HE 1204-0600, HE 2207-0930
and HE 1145-0002 resemble closely the spectrum of HD 5223 in the wavelength 
region 4000 - 5800\AA\,;  they  show  marginally stronger  CN bands
in the wavelength region 5700 - 6800 \AA\,. 

The spectra of HE 0111-1346  and HE 0322-1502 show a very good match with 
the spectrum
of HD 5223, with similar depths in molecular bands and also line depths
of H${\alpha}$,
H$_{\beta}$ and Ba II at 6496 \AA\, appear with similar strength. In 
HE 0322-1502 Na I D appears weakly in emission.

In HE 0151-0341, G-band of CH around 4300 \AA\, and CN band around 4215 \AA\,
have similar strength but C$_{2}$ bands are marginally weaker than in HD 5223.
H${\alpha}$ and H$_{\beta}$ are of equal strength but Na I D and
Ba II at 6496 \AA\ are much weaker than in HD 5223. 
HE 0507-1653 has marginally weaker bands and also Na I D feature is slightly
weaker than in HD 5223; H${\alpha}$,
H$_{\beta}$ and Ba II at 6496 \AA\, appear with similar strength.
 The spectra of HE 0038-0024 and HE 2339-0837 show marginally stronger
CN band around 4215 \AA\, and G-band of CH around 4300 \AA\, but
exhibit slightly weaker C$_{2}$ molecular bands. Na I D feature appears
in weak emission, H${\alpha}$, is of similar strength but Ba II at 6496 \AA\
appear in equal strength in HE 0038-0024 which is marginally weaker in
HE 2339-0837.\\

{\bf HE 1305+0132, HE 1027-2501, HE 1304-2046, HE 0017+055, HE 0319-0215}\\
 The spectra of these stars  also show spectral 
characteristics of CH stars. The spectra  exhibit strong G-band of CH. 
Secondary P-branch head of CH  near 4342 \AA\, is distinctly seen as  
usually seen in CH 
stars spectra. Ca I feature at 4226 \AA\, is weak or undetectable in their 
spectra.   We  place these stars in 
CH group.  Ba II at 4554 \AA\,,  Sr II around 4606 \AA\, and H$_{\beta}$ 
are  seen  in their spectra.
 Strong molecular bands include 
C$_{2}$ Swan bands around 4700 \AA\,  and C$_{2}$ bands around 5165 \AA\, and 
5635 \AA\,. Ba II feature around 6496 \AA\, is blended with contributions 
from CN bands.  CN bands around 5730 \AA\, and 6300 \AA\, are detected. Na I D 
features appear very similar as  seen in most of the CH stars 
except in HE 0319-0215, where this feature appears in weak  emission. 
\\                                                                                
\subsection{Candidate C-N  stars: Description of the spectra} 

{\bf  HE 2319-1534, HE 1008-0636, HE 2331-1329, \\
HE 0207-0211, HE 1107-2105 }\\
The spectra of these stars   show a 
close resemblance with the spectrum of C-N star Z Psc with similar strengths 
of  CN and C$_{2}$ bands in them seen across the wavelength regions. 
In figure 8, we show as an example, a comparison of spectra for three objects
in the wavelength region 5500 - 6800 \AA\, with the spectrum of Z Psc.
 The spectra of HE 2319-1534, HE 1008-0636 and HE 1107-2105
have  low flux below about 4500 \AA\,.  In 
HE 1008-0636 the SiC$_{2}$ bands around 4800 - 5000 \AA\, are seen. These 
red-degraded features are not seen in the other four and  Z Psc.  Na I D 
feature is much deeper in HE 1008-0636 than in HE 2319-1534. 
In the spectrum of HE 2331-1329,  Ca I feature 
at 4226 \AA\, is much weaker than in Z PSc. G-band of CH around
4300 \AA\, and  C$_{2}$  molecular bands are of similar strength
but CN bands are much weaker in HE 2331-1329 than their counterparts in Z PSc.
Na I D, H$_{\alpha}$ and H$_{\beta}$ are marginally detectable in this star.
In HE 0207-0211 CN bands are much weaker than in Z PSc but C$_{2}$ bands
are in good match; Na I D feature is weak and barely detectable.
In HE 0207-0211 and HE 1107-2105 H$_{\alpha}$ and H$_{\beta}$ appear in 
emissionr; these two 
 features   are also seen strongly in emission in the spectra 
of HE 2319-1534 and  HE 1008-0636  
are indicative of a possible strong chromospheric activity or 
shock-waves of the type associated with Mira variables.
  \\

{\bf HE 1228-0402, HE 0915-0327, HE 1254-1130, HE 1501-1500, HE 1259-2601 }\\
The specta of these stars 
 show very low flux below about 4500 \AA\,. Their spectra mostly resemble
the spectra of C-N star  with prominent CN and C$_{2}$ bands seen across the 
wavelength regions.  The spectrum of C-N star V460 Cyg compares closest
to the spectra of these stars.  Na I D feature  is  weaker  in their spectra  
as compared to their  counterparts in V460 Cyg.
 We place these stars in C-N group.  In V460 Cyg the molecular bands of C$_{2}$
as well as CN are much deeper than in Z Psc.  \\

{\bf HE 0002+0053, HE 1104-0957 }\\
 The spectra of these two objects  
greatly resemble the spectrum of  C-R star RV Sct.  C$_{2}$
bands in the spectrum of HE 0002+0053 match closely with  those  in RV Sct 
but CN bands are much weaker. In 
HE 1104-0957 molecular bands due to both  CN and C$_{2}$ are much weaker than
those in RV Sct.  In both the stars,   H$_{\alpha}$ and H$_{\beta}$ are 
weakly seen in absorption. Na I D is marginally detectable but weaker than
in RV Sct. but  G-band of CH around 4300 \AA\, is marginally stronger in these
 stars.  Ca I feature around 4226 \AA\, which appears  weakly in
 the spectrum of RV Sct is missing in  the  spectra of these two stars.
 $^{13}$C isotopic band around 4700\AA\,
is absent in these two stars.  CN
band around 5200 and 5700 \AA\, distinctly seen in RV Sct is marginally
detected in HE 0002-0053  but not seen in HE 1104-0957.
 \begin{table*}
{\footnotesize
{\bf Table 3: Estimated effective temperatures (T$_{eff}$) of the 
candidate CH stars}\\
\begin{tabular}{cccccc}
\hline           

Star Names&T$_{eff}-$ &T$_{eff}-$ &T$_{eff}-$ &$^{12}$C/$^{13}$C &\\
          &  (J-K)   &  (J-H)   &  (V-K)   &        &    \\

\hline 
HE 0017+0055&  3919.1 & 4124.4 &    -   &   1.3  &  \\
HE 0038-0024& 3783.8 &  3929.2 &  4306.2 &   1.9  &    \\
HE 0043-2433& 4263.8 &  4271.3 &  4379.3 &    -   &   \\
HE 0110-0406& 4405.6 & 4444.0 &    -   &    2.1   &  \\ 
HE 0111-1346& 4449.5 &  4481.0 &   -     &   2.5  &   \\
HE 0151-0341& 4619.2 &  4696.5 &  4912.7 &   1.7  &   \\
HE 0308-1612& 4274.8 & 4369.2 &   -    &    2.8  &   \\
HE 0314-0143& 3454.4 & 3561.6 &   -     &  76.8   &   \\ 
HE 0319-0215& 4188.9 & 4314.8 &  4480.4 &   4.7   &   \\ 
HE 0322-1504& 4054.8 &  4293.7 &  4614.6 &   2.2  &   \\
HE 0457-1805& 4097.6 &  4513.4 &  4165.5 &    -   & \\
HE 0507-1653& 4740.2 &  4846.3 &  4983.1 &   6.7  &  \\
HE 0518-2322& 4680.9 &  4689.1 &   -     &    -   &   \\
HE 1027-2501&  -    &   -     &   -      &  1.8   &  \\
HE 1056-1855& 4280.5 &  4427.3 &  --     &    -   &    \\
HE 1145-0002& 3665.5 &  3905.5 &  3691.2 &  1.4   &   \\ 
HE 1125-1357& 3708.6 &  3897.9 &  3910.2 &  3.7   &   \\
HE 1204-0600& 3910.0 &  4102.6 &  3881.6 &  1.7   &   \\
HE 1211-0435& 4710.3 &  4476.0 &  4732.4 &  3.7   &   \\ 
HE 1304-2046& 4073.0 & 4200.9 &  4061.1  &   1.7  &   \\
HE 1305+0132& 3931.4 & 4061.2  &  4131.2  &  1.4  &   \\
HE 1425-2052& 4038.5 &  4179.1 &  3824.0  & 1.7   &    \\
HE 1418+0150 &3781.1 & 4042.8 &    -    &   1.5   &   \\
HE 1429-0551& 4367.4 &  4800.0 &   -    &    1.9  &          \\ 
HE 1446-0112&  3891.9 &  4160.2 &  3854.9 &  1.8  &   \\
HE 1523-1155& 4524.2 &   4491.1 & 4380.8 &   2.5  &      \\ 
HE 1524-0210& 3826.9 &  3942.1  & 4611.2 &  2.3   &  \\
HE 1528-0409& 4666.7 &  4662.4 &  4388.4 &   2.4   &     \\
HE 2144-1832&  3922.2 &  4226.4 &   -     &  2.1   &  \\ 
HE 2145-1715& 4019.1 & 3862.1 &   4214.1 &   2.2  &  \\  
HE 2207-0930& 3628.5 &  3751.7 &  3736.5 &   1.4  &  \\ 
HE 2207-1746& 4378.8 &  4448.4 &   -     &   3.2  &   \\ 
HE 2218+0127& 5544.6 &  5631.3 &   -     &   4.2  &  \\
HE 2221-0453& 4231.7 &  4487.1 &  4163.8 &  12.6  &    \\
HE 2339-0837& 4592.6 &  4499.1 &  5104.5 &   2.3  &   \\
\hline 
HD 26      &         &        &        &    5.9  &   \\
HD 209621  &         &        &        &   8.8  &  \\
HD 5223    &         &        &        &   6.1  &  \\
\hline 

\end{tabular}

}
\end{table*}

\section{Atmospheres of CH stars }

\subsection{ Effective temperature}

Preliminary estimates of the  effective temperatures of the candidate  
CH stars are determined by using 
 temperature calibrations  derived by Alonso et al. (1996).
These calibrations were derived by using a large 
number of lower main sequence stars and subgiants, whose temperatures were 
measured by infrared flux method, and holds within a temperature and 
metallicity range 4000 ${\le}$ T$_{eff}$ ${\le}$ 7000 K and 
-2.5 ${\le}$ [Fe/H] ${\le}$ 0 . 
This calibration relates T$_{eff}$ with Stromgren indices as well
as [Fe/H] and colours (V-B), (V-K), (J-H) and (J-K).  
By considering the  uncertainties arising from different sources such as 
uncertainties in the Stromgren photometry, reddening and the calibration 
of the absolute flux in the infrared, Alonso et al. (1996) estimated
an uncertainty of ${\sim}$ 90 K in T$_{eff}$ determination.
The broad band B-V colour is often used for the determination of T$_{eff}$,
however B-V colour of a giant star depends not only on T$_{eff}$ but also
on metallicity of the star and the molecular carbon absorption features,
due to the effect of CH molecular absorption in the B band. For this reason,
we have not used the empirical T$_{eff}$ scale for the B-V colour indices.
Since there is a negligible difference between the 2MASS infrared photometric
system and the photometry data measured on TCS system used by Alonso et al.
(1998) in deriving the T$_{eff}$ scales; we have used the empirical
T$_{eff}$ scales with 2MASS photometric data. 
We have further assumed that the effects of reddening on the measured colours
are  negligible.  In estimating the
T$_{eff}$ from T$_{eff}$ - (J-H)  and T$_{eff}$ - (V-K) relations 
 we had to adopt a value for metallicity
of the star as the metallicity of these stars are not known. We assumed
the metallicity of the stars to be same as their closest comparison star.
This assumption has definitely affected the accuracy of the T$_{eff}$ 
measurements. Estimated effective temperatures are listed in Table 3.

For a reliable determination of metallicity, effective  temperatures
and  chemical compositions of these stars,  observation  at high resolution 
is necessary.  High resolution spectra will also enable us  for an accurate 
measurement of $^{12}$C/$^{13}$C ratios. 

\subsection{ Isotopic ratio $^{12}$C/$^{13}$C from molecular band depths}

Carbon isotopic ratio $^{12}$C/$^{13}$C  provides   an important
 probe of stellar evolution but low resoltuion of the spectra
does not allow a meaningful estimation of this ratio.

We have estimated the ratio of the molecular band depths using 
the bands of 
(1,0) $^{12}$C$^{12}$C ${\lambda}$4737 and
(1,0) $^{12}$C$^{13}$C ${\lambda}$4744.
For a majority of the sample stars,
we find from the depths of molecular bands the ratio 
 $^{12}$C/$^{13}$C $\sim$ 3,
with an exception of three  stars for which this ratio is  7, 13 
and 77 respectively.
The ratios   are presented  in Table 3.
This ratio measured on  the spectra of the welknown CH stars 
HD 26, HD 5223 and HD 209621 are respectively
5.9 , 6.1  and 8.8 .  Tsuji et al. (1991) had  suggested  two
kinds of CH stars; one with very high $^{12}$C/$^{13}$C ratio and the other 
with  the values less than about 10. 
Our estimated ratios of  $^{12}$C/$^{13}$C are consistent with this.

Several explanations on the significance of the range of values 
 of $^{12}$C/$^{13}$C  ratios are put forward in terms of the stars
evolutionary scenarios. 
One explanation for a lower value of $^{12}$C/$^{13}$C ratio is that,
generally, the $^{12}$C/$^{13}$C ratio and total carbon abundances decrease 
due to the convection which dredges up the products of internal CNO cycle to 
stellar atmosphere as ascending RGB. If it reaches AGB stage,  fresh $^{12}$C 
may be supplied from the internal He burning layer to stellar surface leading 
to an increase of  $^{12}$C/$^{13}$C ratio  again. Since the abundance 
anomalies observed in  CH giants are believed to have originated by the 
transfer of mass from a now extinct AGB companion, the CH giant's atmosphere 
should be enhanced in triple $\alpha$ products from the AGB star's 
interior- primarily $^{12}$C.  This explanation is in favour of  stars which 
give high $^{12}$C/$^{13}$C ratios.  The low carbon isotope ratios imply that 
the material transferred from the now unseen companion has been mixed into 
the CN burning region of the CH star or constitutes a minor fraction of the 
envelope mass of the CH star, thus giving isotope ratios typical of stars 
on their first ascent of the giant branch.

\section{Concluding Remarks}
Large samples of high latitude carbon stars such as one reported by 
Christlieb et al. allows a search for different kinds of carbon stars; 
the present work is
a step in this direction. The sample of carbon star candidates offered
by Christlieb et al. being high latutude objects, smaller initial masses
and possible lower metallicity, it is likely that a reasonable fraction 
of it could be CH stars. Indentification of several  CH stars and description 
of their spectra are the main results of this paper.
 Another effort is known to be underway to make a medium-resolution
spectroscopic study of the  complete sample of stars from Christlieb
 et al. 2001 (Marsteller et al. 2003, Beers et al. 2003).
 From the sample list we have acquired spectra for  ninety one
stars in the first phase of observation. Out of these,
fifty one  objects were found to  exhibit  strong C$_{2}$ molecular bands 
 in their spectrs of which thirteen  stars
have low flux below about 4300 \AA\,. 
Twenty five objects show weak or moderate CH and CN bands, 
twelve  objects show weak but detectable CH bands in their spectra and
there are three objects which do not show any molecular bands due to C$_{2}$,
CN or CH in their spectra.
As an example, in figure 9  we show   
three  spectra:  a  spectrum of 
HE 0443-1847 which exhibits very weak molecular bands due to CN around
4215 \AA\, and a weak G-band of CH around 4300 \AA\, (but no C$_{2}$
molecular bands); a  spectrum of HE 0930-0018 which 
 show   a  weak signature of G-band of CH around 4300 \AA\,
and a  spectrum of HE 1227-3103 which do not show any molecular bands
due to C$_{2}$, CN or CH in its spectrum.

Although spectroscopically, appearance of strong C$_{2}$ molecular bands 
is an obvious
indication of a  star being a carbon star, the conventional defination of a
carbon star is a star with C/O ${\ge}$ 1 (Wallerstein et al. 1997). Hence if
one adopts this conventional definition non appearance of any C$_{2}$ molecular
bands will not necessarily disqualify a star from being a carbon star as this
does not exclude the condition C/O ${\ge}$ 1; which at our resolution of
the spectra is not derivable.

Westerlund et al. (1995) defined dwarf carbon stars as having J-H ${\le}$ 0.75,
 H-K ${\ge}$ 0.25 mag. None of the  stars occupies a region defined
by these limits in J-H, H-K plane. With respect to J-H, H-K colours
there is a clear separation between the C-N type stars and dwarf carbon-star
populations; there are CH stars with J-H ${\le}$ 0.75 but their H-K values
are less than the lower limit of 0.25 mag set for dwarf carbon stars.
We find that the sample of stars under investigation is comprised mostly of 
CH stars and a small number of C-N and C-R  stars.

We have derived the effective temperatures  of the candidate CH stars 
 from correlations of Alonso et al. (1996) making use
of (J-K), (J-H) and (V-K) colour indices. They vary over a wide range of 
temperature with an average of $\pm$ 240 K.
These temperature estimates provide a preliminary temperature check for the 
program stars and can   be used  as starting values in deriving atmospheric 
parameters from high resolution spectra using model atmospheres. For majority 
of the sample stars, we find   carbon isotopic ratio 
$^{12}$C/$^{13}$C $\sim$ 3 with an exception of three stars HE 0507-1653,
HE 2221-0453 and HE 0314-0143 for which this ratio is 7, 13 and 77 
respectively.  
It was suggested  by Tsuji et al. (1991) that there could be two
kinds of CH stars, one with very high $^{12}$C/$^{13}$C ratio and the
other with values ${\sim}$ 10. Our $^{12}$C/$^{13}$C estimates are consistent
with this.
This range of ratios is  the same as found for the population II giants and 
globular cluster giant stars (Vanture  1992). Different evolutionary 
scenarios are held responsible for the two groups of CH stars, one  with 
high  and the other with low $^{12}$C/$^{13}$C ratios.
 
From radial velocity survey   CH stars are known  to be binaries. 
For the moderately metal-poor classical CH stars ([Fe/H] ${\sim}$ -1.5),
a  scenario for abundance anomalies and the origin of carbon was  proposed
in which the carbon-enhanced star is a member of a wide binary system that
accreted material from a former primary, during the asymptotic giant branch
(AGB) phase of the latter, as described by McClure \& Woodsworth (1990).
In such a scenario  CH stars with large $^{12}$C/$^{13}$C ratios indicates 
that their atmosphere is  enhanced in triple $\alpha$ products.
The process of convection dredges up the products of internal CNO cycle to
the stellar atmospheres as ascending RGB and this leads to a decrease of
or a small value of $^{12}$C/$^{13}$C ratio and a small total carbon
abundance;  on reaching the AGB stage $^{12}$C/$^{13}$C
ratio increases again  due to the receipt of fresh $^{12}$C supplied from the
internal helium burning layer to the stellar surface.
According to the models of  McClure (1983, 1984)  and   McClure \& Woodsworth 
(1990) the CH binaries have orbital characteristics consistent with the
presence of a white dwarf companion,  these stars have conserved the products 
of carbon rich primary and survived untill the present in the Galactic halo.

However, in  case of a few 
 carbon-enhanced, metal-poor stars  (subgiants)  monitoring of radial 
velocity over a period of eight years did not reveal radial velocity
variations greater than 0.4 km s$^{-1}$ which  is against the mass transfer 
scenario for these  stars (Norris et al. 1997a, Aoki et al. 2000, Preston
and Sneden 2001). 
 Furthermore, it is expected that
the star we observe today should display an enrichment of $s$-process
elements, produced by the former primary in its AGB phase, while the
carbon-enhanced metal-poor  star CS 22957-027 (Norris et al. 1997b, 
Bonifacio et al. 1998),
as well as the  stars reported  by Aoki et al. 2000
do not exhibit this behaviour.
The carbon-enhanced metal-poor  stars that do show $s$-process enrichment 
 provide  strong observational constraints for theoretical
models of the structure, evolution and nucleosynthesis of early-epoch
AGB stars and permit studies of the $s$-process operating at very low
metallicities. 
It was shown by Goriely \& Siess (2001) that even at the absence of iron
seeds efficient production of $s$-process elements can take place at
zero metallicity provided protons are mixed into carbon-rich layers
producing $^{13}$C, which acts as a strong neutron source via 
 $^{13}$C(${\alpha}$ ,n)$^{16}$O. The recent discovery of
carbon-enhanced metal-poor  stars with strong overabundances of Pb 
support these predictions ( Aoki et al. 2000, Van Eck et al. 2001). 
Thus CH stars being  the most prominent of the few types of heavy 
element stars  that exist
in both the field of the Galaxy and globular clusters  are an important
class of objects which can provide 
some of the very few direct observational tests to stellar evolution
theory.  \\

 While in the present work,  the spectra of the stars
listed in table 1, are discussed  the analysis and description  of the  
spectra of the
stars listed in table 2 will be discussed in a  subsequent work.\\

 {\it Acknowledgement}\\
 We thank  the staff at IAO and at the remote control station  at CREST,
Hosakote for assistance during the observations.  This work made use of 
the SIMBAD astronomical database, operated at CDS, Strasbourg, Franch, and 
the NASA ADS, USA. I am grateful to the referee Prof Timothy Beers for
his many constructive suggestions which has improved  considerably the
readability of  the paper. The author would also like to thank Professor
N. K. Rao for his guidance in the observational program, and helpful 
suggestions.\\

{}

\begin{figure*}
\epsfxsize=18truecm
\epsffile{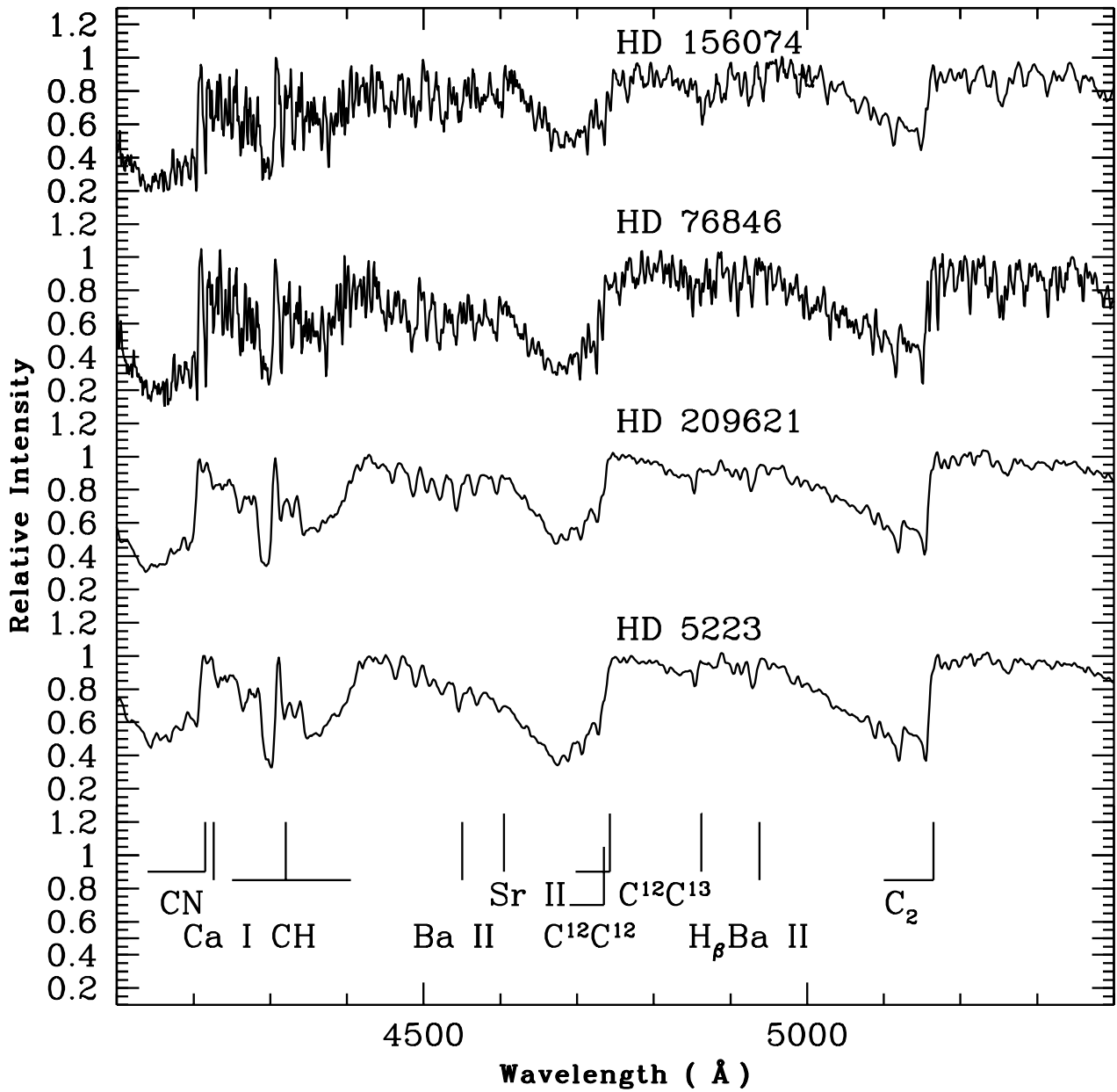}
\caption{
A comparison of the  spectra of a pair of  C-R stars HD 156074 and HD 76846 
and a pair of  CH stars HD 5223 and HD 209621 in the wavelength region
4100-5400 \AA\,. The most prominent features noticeable are marked on 
the figure. \label{fig 2}} 
\end{figure*}

\begin{figure*}
\epsfxsize=18truecm
\epsffile{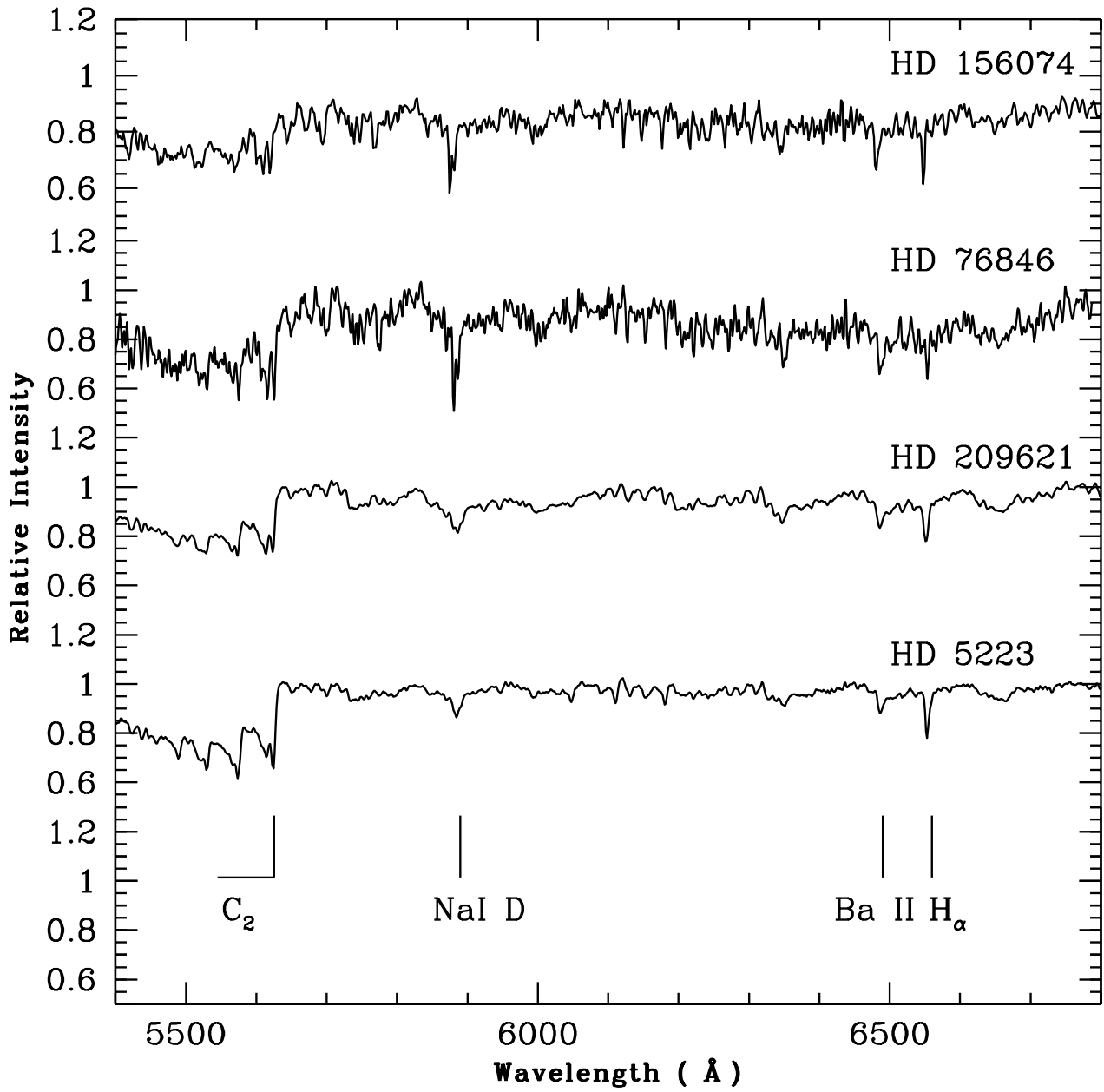}
\caption{
Same as figure 2 but for the wavelength region 5400-6800 \AA\,. The 
prominent features of Na I D, Ba II at 6496 \AA\, H$_{\alpha}$  and 
C$_{2}$ molecular bands around 5635 \AA\, are
indicated. \label{fig 3}} 
\end{figure*}
\clearpage

\begin{figure*}
\epsfxsize=18truecm
\epsffile{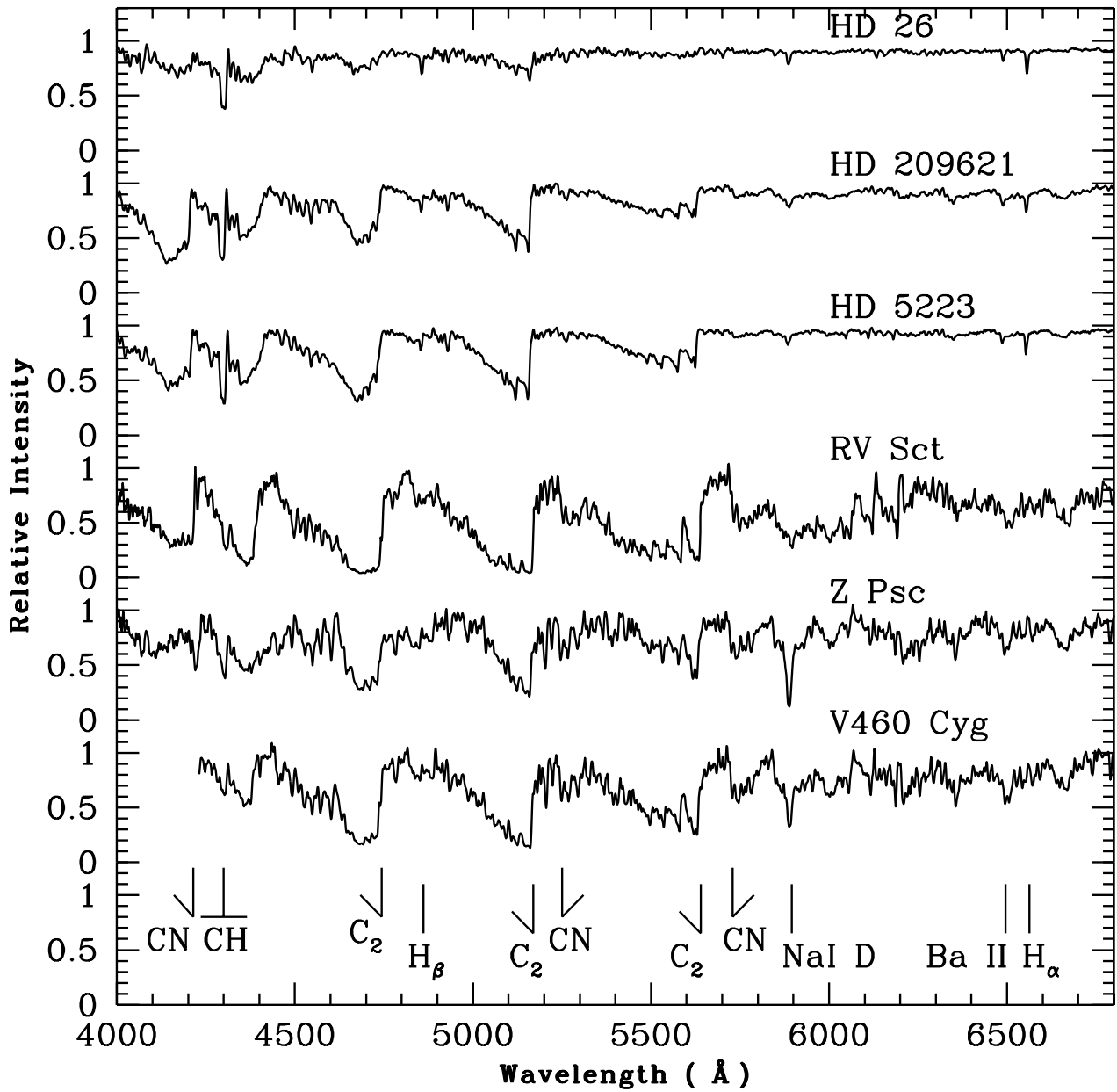}
\caption{
 The spectra  of the comparison stars 
  in the wavelength region
4000-6800 \AA\,.    \label{fig 4}} 
\end{figure*}
\clearpage

\begin{figure*}
\epsfxsize=18truecm
\epsffile{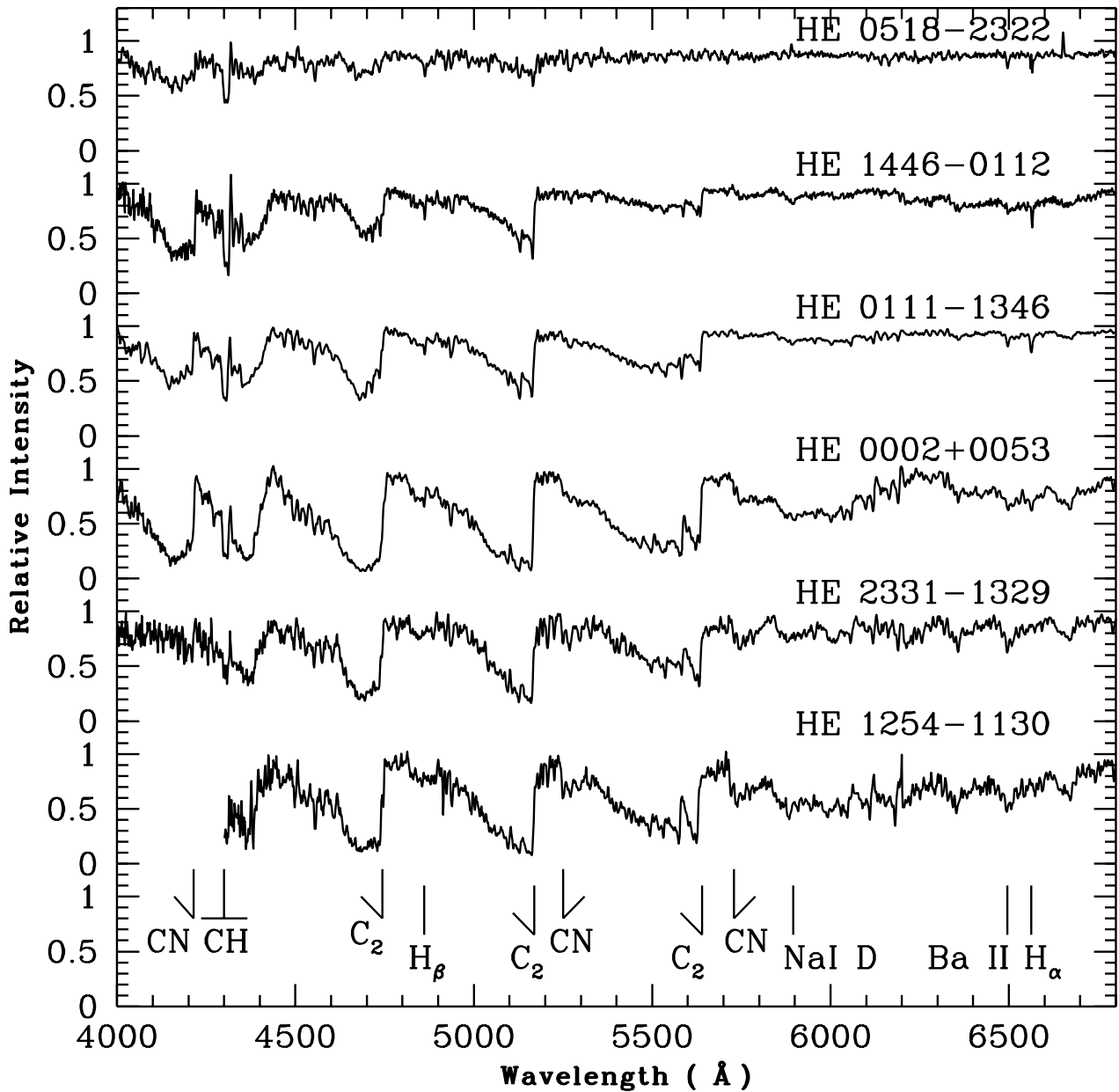}
\caption{
 The figure shows one example  from the HE stars corresponding to the 
comparison stars presented in figure 4, in the  top to 
bottom sequence,   in the  wavelength region 4000-6800 \AA\,. 
The locations of some prominent features  seen in the spectra 
are marked on the figure.
HE 1254-1130 has low flux below about 4400 \AA\,. Ba II at 6496 \AA\,
and H$_{\alpha}$ seen in the top three stars spectra are not
detectable in the lower three stars spectra. Except for the Na I D
feature which is barely detectable in the spectra of HE 2331-1329 and
HE 1254-1130, these  two stars spectra resemble closely  to their 
comparison stars spectra of  Z Psc and V460 Cyg respectively.
 \label{fig 5}} 
\end{figure*}
\clearpage

\begin{figure*}
\epsfxsize=18truecm
\epsffile{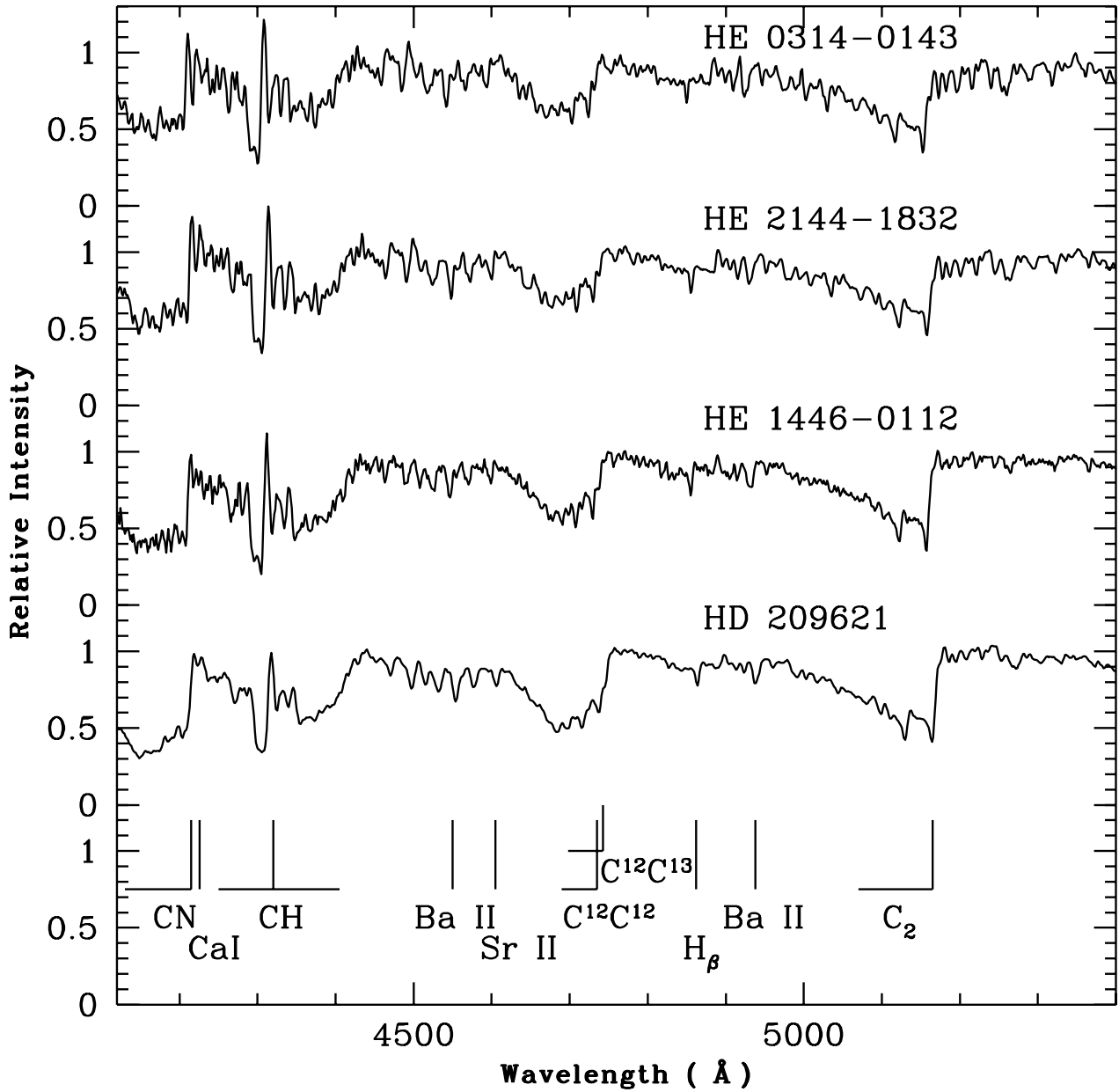}
\caption{
  The figure shows  a comparison of  three  HE stars   spectra in the 
wavelength region  4120 - 5400 \AA\,  with the  comparison  star's spectrum 
of HD 209621. Prominent features seen in the spectra are marked on the figure. 
\label{fig 6}} 
\end{figure*}
\clearpage

\begin{figure*}
\epsfxsize=18truecm
\epsffile{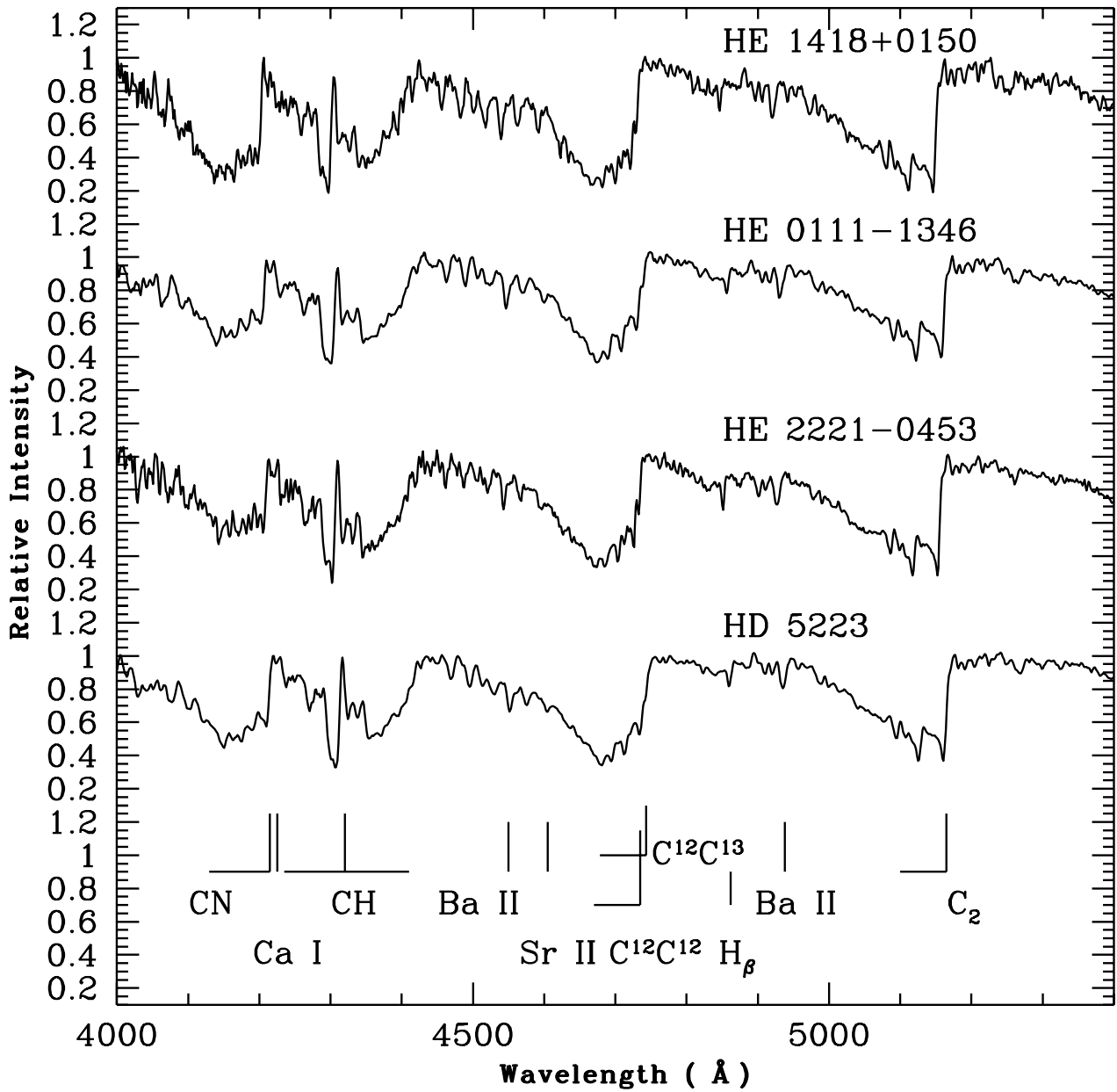}
\caption{
 The  figure shows  a comparison of  three  HE stars  spectra in the 
wavelength region  4000 - 5400 \AA\,  with the  comparison star's spectrum 
of HD 5223. Some of the prominent features seen in the spectra are marked 
on the figure.
 \label{fig 7}} 
\end{figure*}
\clearpage

\begin{figure*}
\epsfxsize=18truecm
\epsffile{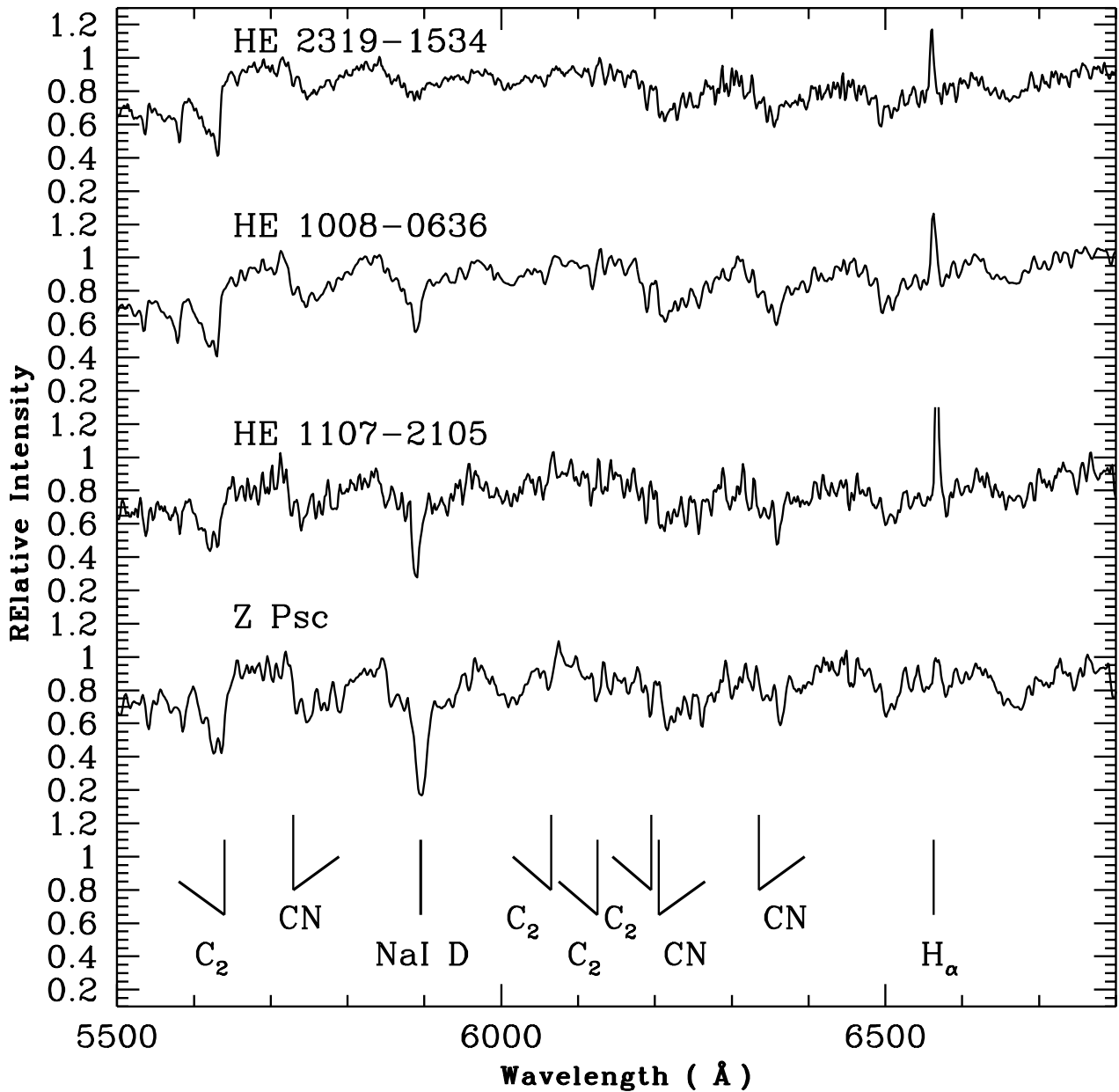}
\caption{
 A comparison of the  spectra of the  candidate C-N stars HE 2319-1534,  
HE 1008-0636 and HE 1107-2105 with the spectrum of Z Psc
in the wavelength region 5500 \AA\, to 6800 \AA\,. 
The bandheads of the prominent molecular bands,  NaI D and 
H$_{\alpha}$ are marked on the figure. H$_{\alpha}$ is seen strongly in 
emission in the HE stars spectra.
 \label{fig 8}} 
\end{figure*}
\clearpage

\begin{figure*}
\epsfxsize=18truecm
\epsffile{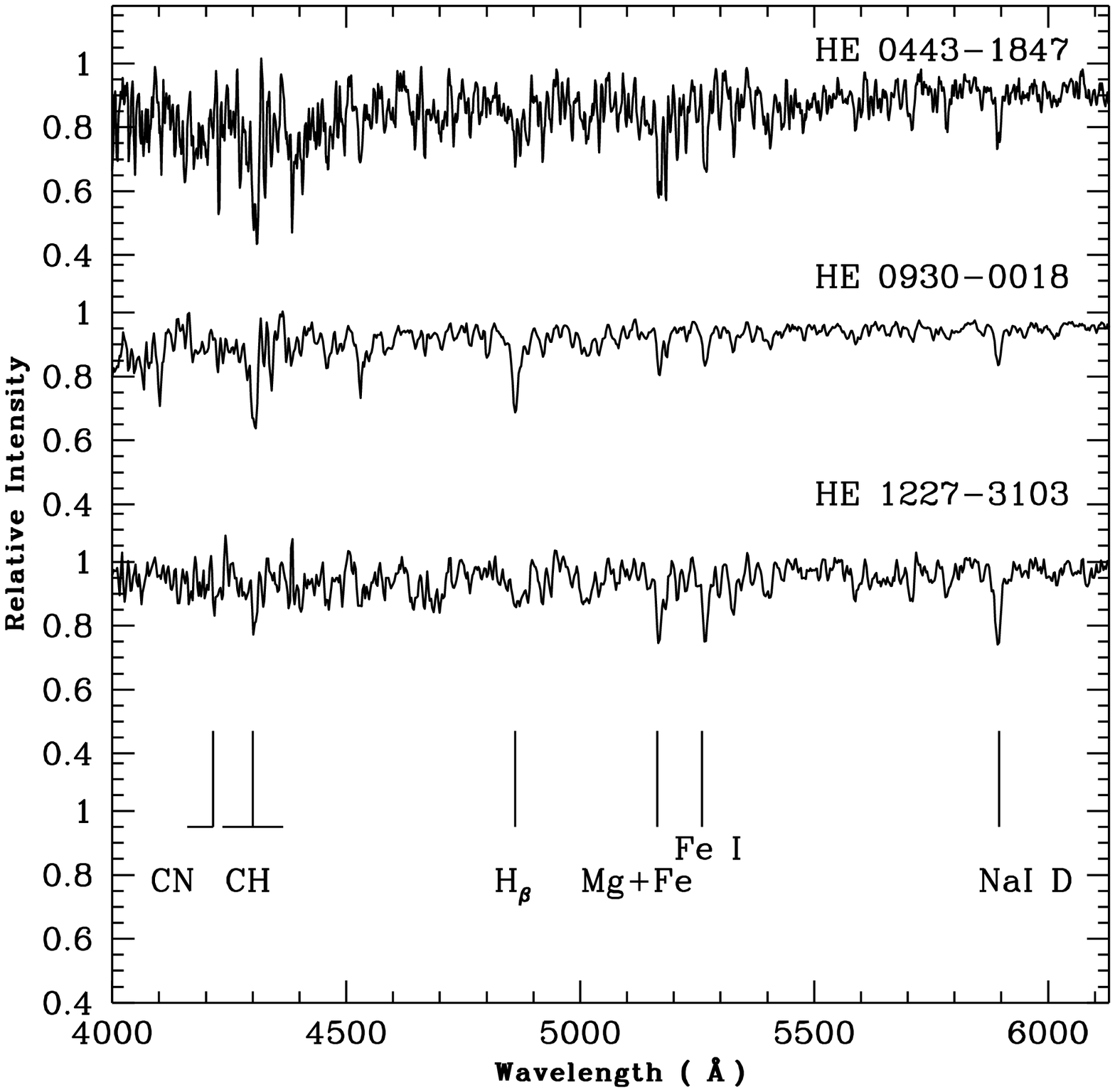}
\caption{
This figure   demonstrates  three examples of HE stars 
in the wavelength range 4000 - 6130 \AA\,.   
The   spectrum of 
HE 0443-1847 which exhibits very weak molecular bands due to CN around
4215 \AA\, and a weak G-band of CH around 4300 \AA\, (but no C$_{2}$
molecular bands); the   spectrum of HE 0930-0018 which 
 show   a  weak signature of G-band of CH around 4300 \AA\,
and the  spectrum of HE 1227-3103 which does not show  presence of any 
molecular bands due to C$_{2}$, CN or CH in its spectrum.
 \label{fig 9}} 
\end{figure*}
\clearpage


\begin{thebibliography}{}
\bibitem {}  Alonso, A. et al. 1996, A\&A, 313, 873
\bibitem {}  Alonso, A. et al. 1998, A\&AS, 131, 209
\bibitem {}  Aoki, W. \& Tsuji, T. 1997, A\&A, 317, 845
 \bibitem {}  Aoki, W., Norris, J. E., Ryan, S. G., Beers, T. C. \& Ando, H.
2002, ApJ, 567, 1166
 \bibitem {}  Aoki, W., Norris, J. E., Ryan, S. G., Beers, T. C., \& Ando, H.
  2000, ApJ, 536,  L97
{\it The Third Stromlo Symposium: the Galactic Halo, ed. B. K. Gibson, T. S.
Axelrod \& M. E. Putman } (San Francisco: ASP), 202
\bibitem {}  Barnbaum, C., Stone, R. P. S. \& Keenan, P. 1996, ApJS, 105, 419
\bibitem {} Beers, T. C. et al. 2003, IAUJD..15E..59B
\bibitem {} Beers, T. C., Preston, G. W. \& Shectman, S. A. 1992, AJ, 103, 1987
\bibitem {} Beers, T. C. 1999, in ASP Conf Ser. 165,
\bibitem {} Bonifacio, P. et al. 1998, A\&A, 332, 672
\bibitem {}  Chriestlieb, N. et al. 2001, A\&A, 375, 366
\bibitem {}  Dominy, J. F. 1984, ApJS, 55, 27
\bibitem {} Goriely, S \& Siess, L. 2001, A\&A, 378, L25
\bibitem {}  Green, P. J. et al. 1992, ApJ, 400, 659
\bibitem {}  Green, P. J. et al. 1994, ApJ, 433, 319
\bibitem {}  Green, P. J. \& Margon, B. 1994,  ApJ, 423, 723 
\bibitem {}  Harding, G. A., 1962, Observatory, 82, 205
\bibitem {}  Hartwick, F. D. A. \& Cowley, A. P. 1985, AJ, 90, 2244
\bibitem {} Hill, V. et al. 2002, A\&A, 387, 560
\bibitem {}  Lambert, D. L. et al. 1986, ApJS, 62, 373
\bibitem {} Marsteller, B. et al.  2003, AAS...20311216M20311216M
\bibitem {}  McClure, R. D. \& Woodsworth, A. W., 1990, ApJ, 352, 709
\bibitem {}  McClure, R. D. 1983, ApJ, 268, 264
\bibitem {}  McClure, R. D. 1984, ApJ, 280, L31
\bibitem {}  Norris, J. E., Ryan, S. G., \& Beers, T. C. 1997a, ApJ, 488, 350
\bibitem {}  Norris, J. E., Ryan, S. G., \& Beers, T. C. 1997b, ApJ, 489, L169
\bibitem {}  Norris, J. E., Ryan, S. G., Beers, T. C., Aoki, W. \&  Ando, H.
2002, ApJ, 569, L107
\bibitem {} Preston, G. W. \& Sneden, C 2001, AJ, 122, 1545
\bibitem {} Rossi, S., Beers, T. C. \& Sneden, C. 1999, in ASP Conf Ser. 165,
{\it The Third Stromlo Symposium: the Galactic Halo, ed. B. K. Gibson, T. S.
Axelrod \& M. E. Putman } (San Francisco: ASP), 264
\bibitem {}  Tsuji, T. et al. 1991, A\&A, 252, L1
\bibitem {}  Totten, E. J. \& Irwin, M. J. 1998, MNRAS, 294, 1
\bibitem {}  Totten, E. J. et al. 2000, MNRAS, 314, 630
\bibitem {} Van Eck, S., Goriely, S.i, Jorissen, A., \& Plez, B. 2001, Nature,
412, 793
\bibitem {}  Vanture, Andrew D. 1992, AJ, 104, 1997
\bibitem {}  Wallerstein, G. 1969, ApJ, 158, 607
\bibitem {}  Wallerstein, G. \& Knapp, G. 1998, ARA\&A, 36, 369 
\bibitem {}  Westerlund, B. E., Azzopardi, M., Breysacher, J. 
Rebeirot, E. 1995, A\&A, 303, 107
\bibitem {}  Yamashita, Y. 1975, PASJ, 27, 325
\bibitem {}  Zinn, R. 1985, ApJ, 293, 424
\end{thebibliography}
\end{document}